\documentclass[aps,pra,twocolumn,showpacs,superscriptaddress]{revtex4}  
\usepackage{amsmath,amssymb,amstext,graphicx,bm}
\usepackage{color}
\usepackage{pifont}
\makeatletter

\begin{document}
\title{Dynamic response of 1D bosons in a trap}

\author{Vitaly N. Golovach} 
\affiliation{Physics Department, Arnold
  Sommerfeld Center for Theoretical Physics, and Center for
  NanoScience, Ludwig-Maximilians-Universit\"at, 80333 Munich,
  Germany}

\author{Anna Minguzzi} 
\affiliation{Universit\'{e} Joseph Fourier,
  Laboratoire de Physique et de Mod\'{e}lisation des Milieux
  Condens\'{e}s, C.N.R.S. B.P. 166, 38042 Grenoble, France}

\author{Leonid I. Glazman} 
\affiliation{Department of Physics, Yale
  University, New Haven, Connecticut, USA, 06520}

\date{\today}

\begin{abstract} 
We calculate the dynamic structure factor $S(q,\omega)$ of a
one-dimensional (1D) interacting Bose gas confined in a harmonic trap. The
effective interaction depends on the strength of the confinement enforcing the
1D motion of atoms; interaction may be further enhanced by superimposing an
optical lattice on the trap potential. In the compressible state, we find that
the smooth variation of the gas density around the trap center leads to
softening of the singular behavior of $S(q,\omega)$ at Lieb-$1$ mode compared
to the behavior predicted for homogeneous 1D systems. Nevertheless, the
density-averaged response $\bar{S}(q,\omega)$ remains a non-analytic function
of $q$ and $\omega$ at Lieb-$1$ mode in the limit of weak trap confinement. The
exponent of the power-law non-analyticity is modified due to the inhomogeneity
in a universal way, and thus, bears unambiguously the information about the
(homogeneous) Lieb-Liniger model. A strong optical lattice causes formation of
Mott phases.  Deep in the Mott regime, we predict a semi-circular peak in
$S(q,\omega)$ centered at the on-site repulsion energy, $\omega=U$. Similar
peaks of smaller amplitudes exist at multiples of $U$ as well. We explain the
suppression of the dynamic response with entering into the Mott regime,
observed recently by D. Cl\'{e}ment et al., Phys. Rev. Lett. {\bf 102}, 155301
(2009), based on an $f$-sum rule for the Bose-Hubbard model.  
\end{abstract} 

\maketitle

\section{Introduction}  

Experiments with ensembles of trapped cold atoms rejuvenated a number of
directions in quantum many-body physics, posing new problems within that
seemingly well-established field. One of the directions deals with the
properties of interacting bosons confined to a strongly anisotropic,
``one-dimensional'' (1D) trap. The effective repulsion between bosons is
enhanced by making the trap narrower\cite{Weiss} or by imposing a periodic
potential, thus constraining the kinetic energy of bosons moving along the
trap.\cite{Bloch}  

The increase of the effective interparticle interaction affects the static and
dynamic characteristics of the 1D Bose system.  Modification of the
one-particle momentum distribution in the case of interaction enhanced by
periodic potential was observed in Ref.~\onlinecite{Bloch}.  A similar
experiment without a periodic potential but in a tighter 1D trap was reported
in Ref.~\onlinecite{Weiss}.  The experimental data is in a reasonable agreement
with the predictions of the integrable Lieb-Liniger model.\cite{LiebLiniger}
That model allows one to find quantitatively the single-particle distribution
function of 1D bosons at any interaction strength in the absence of optical
lattice, and establish the qualitative features of the distribution in the
presence of the lattice, at least in the limit of small particle density.

The dynamic response of 1D interacting bosons can not be derived directly from
the thermodynamic Bethe Ansatz solution~\cite{LiebLiniger} of the Lieb-Liniger
model. Nevertheless, a significant progress has been made in finding the
universal singularities of the dynamic response
analytically~\cite{Khodas,Imambekov1,Imambekov2,ChernyBrand} and determining
the general structure of the response in a broader range of energies
numerically,\cite{Caux} utilizing the ideas of algebraic Bethe Ansatz.  The
most suitable experimental method for investigating the dynamic structure
factor $S(q,\omega)$ is Bragg
spectroscopy.\cite{Ketterle,Steinhauer1,Steinhauer2,Richard,Muniz,Papp,Inguscio1,Inguscio2}
By a nonlinear mixing of two laser beams, it allows for an independent control
of the wave vector $q$ and frequency $\omega$ of the perturbation.  Recent
experiments\cite{Inguscio1,Inguscio2} have demonstrated the use of this method
in 1D systems to study effects of interaction and periodic confinement.

Possible complications in comparing experimental data with theory arise from
deviations of a real atomic system from the ideal Lieb-Liniger model.  Even in
most favorable cases it is impossible to avoid a ``soft'' confining potential
applied along the 1D trap, which makes the Bose liquid inhomogeneous.  An
optical lattice potential, applied in addition to the trap confinement to
enhance the effective interaction between particles, brings in additional
complications.  In this paper, we develop the theory of dynamic response for an
inhomogeneous 1D boson system confined by a ``soft'' trap potential, both in
the presence and in the absence of an optical lattice.  In the absence of
optical lattice, the system is a compressible inhomogeneous Bose liquid.  The
soft confinement modifies the dynamic response as compared to the homogeneous
1D system.  For weak interaction, the density-averaged response
$\bar{S}(q,\omega)$ of the liquid has an asymmetric peak as a function of
frequency $\omega$ at the Lieb-$1$ mode $\varepsilon_+(q)$, evaluated at the
maximal gas density in the trap.  The general shape of $\bar{S}(q,\omega)$ is
skewed towards low frequencies and the peak shape is a power-law singularity,
with an exponent dependent on the interaction strength.  For stronger
interaction, the maximum shifts below $\varepsilon_+(q)$.

In the presence of an optical lattice, the atomic cloud may become a mixture of
incompressible Mott phases in equilibrium with Bose liquid.  The tight-binding
limit of the lattice is described by the Bose-Hubbard model with an on-site
repulsion energy $U$ and an inter-site hopping matrix element $J$.  The dynamic
response of a Mott phase is unaffected by the trap potential and is centered at
frequency $\omega\approx U$.  The peak has a semicircular shape, a
$q$-dependent width, and a spectral weight $\propto \left(J/U\right)^2$.
Unlike the Mott phase, the Bose liquid phase has dynamic response at small
frequencies $\omega\sim J$, and this response does not vanish in the limit
$U\to\infty$.  Therefore, in a mixed state of incompressible and compressible
liquids the largest weight in the dynamic response comes from the compressible
liquid and is at frequencies $\omega\sim J$.  We analyze in greater detail the
simplest case, in which one Mott phase (with site occupancy $1$) extends nearly
over the entire cloud, ending with compressible-liquid caps on both ends.

The paper is organized as follows.  In Sec.~\ref{SecQualCons}, we give a
qualitative description of the main effects considered in the paper.  In
Sec.~\ref{DSFofHMPh}, we calculate the dynamic structure factor of the
homogeneous Mott phase.  In Sec.~\ref{SecAverDynStrFactor}, we find dynamic
responses averaged over the trap inhomogeneity.  Attention is given to the
inhomogeneous Bose liquid state as well as to a mixed state of coexisting Mott
and Bose liquid phases.  In Sec.~\ref{SecDSFandLA}, we discuss ways to measure
the dynamic structure factor, the $f$-sum rules in the presence of an optical
lattice, and recent experiments on dynamic response of trapped atomic systems.
Appendix~\ref{AppFiniteSizeEffects} deals with the finite-size effects due to
the trap confinement.  In Appendix~\ref{AppAvOverTubes}, we study the effects
due to averaging over an ensemble of 1D systems.  In
Appendix~\ref{AppDynResFinTemp}, we consider dynamic response at finite
temperatures.

\section{Qualitative consideration and main results} 
\label{SecQualCons}

Due to the confining potential, the density of atoms varies along the trap. If
in addition to the potential of the trap an optical lattice is created, then
the variation of the density may lead to coexistence of compressible and
incompressible phases in the trap. In this section, we first identify the
domain of parameters where the inhomogeneous cold-atom system consists of no
more than two phases.  Next, we summarize our main results for the dynamic
structure factor averaged over the trap at zero temperature.

\subsection{Phases of cold atoms subject to the trap and optical
  lattice potentials}
\label{SecPhases}

The presence of an optical lattice changes the dispersion relation of
individual bosons.  The effect of interaction becomes stronger with the
narrowing of the single-particle bands.  We consider here the limit of strong
lattice potential, in which narrow bands are separated by wide gaps.
Furthermore, we assume that only the lowest band is populated and thus
concentrate on a one-dimensional Bose-Hubbard model with inter-site hopping
matrix element $J$ and on-site repulsion energy $U$,
\begin{eqnarray}
H_{\rm BH}=&\!-&\!\!J\sum_l\left(b_{l+1}^\dagger b_l+b_{l}^\dagger
  b_{l+1}\right)
+\frac{U}{2}\sum_l b_l^\dagger b_l\left(b_l^\dagger b_l-1\right)\nonumber\\
&+&\sum_l [V(al)-\mu]b_l^\dagger b_l,
\label{BH}\\
&& V(x)=\epsilon_0\left(\frac{x}{a}\right)^2,\quad l=0,\pm 1,\pm 2,\dots\,.
\label{conf}
\end{eqnarray}
Here $V(x)$ is the confinement potential, $a$ is the optical lattice constant,
and $\mu$ is the chemical potential, which depends on the total number $N$ of
atoms in the trap. 

In the following, we view $N$ and $J$ as control parameters ($J$ depends
exponentially on the amplitude of the optical lattice potential). It is
convenient to introduce the characteristic scale,
\begin{equation}
N_0=\sqrt{\frac{U}{\epsilon_0}},
\label{N0}
\end{equation}
and measure $N$ in units of $N_0$, and $J$ in units of $U$.  The meaning of
$N_0$ becomes clear from considering the $J=0$ limit: $2N_0$ is the largest
number of atoms fitting in the trap without causing double occupancy of any
lattice site. In some of the existing experiments with $^{87}$Rb
atoms,~\cite{Bloch,Inguscio1} the harmonic trap frequency was typically
$2\pi\times 60$~Hz, the lattice parameter $a\approx 415$~nm, and interaction
$U/(2\pi)\sim 0.13-2$~kHz which corresponds to the value of $N_0$ in the
range from $7$ to $30$.  For investigating the dynamics of the compressible
phase, it is desirable to raise the value of $N_0$, which may be achieved by
reducing the trap frequency. In the typical conditions of an experiment with
multiple 1D traps,~\cite{Bloch} this would allow to place more atoms in each of
the traps.

\begin{figure}[t]
\includegraphics[width=0.95\columnwidth]{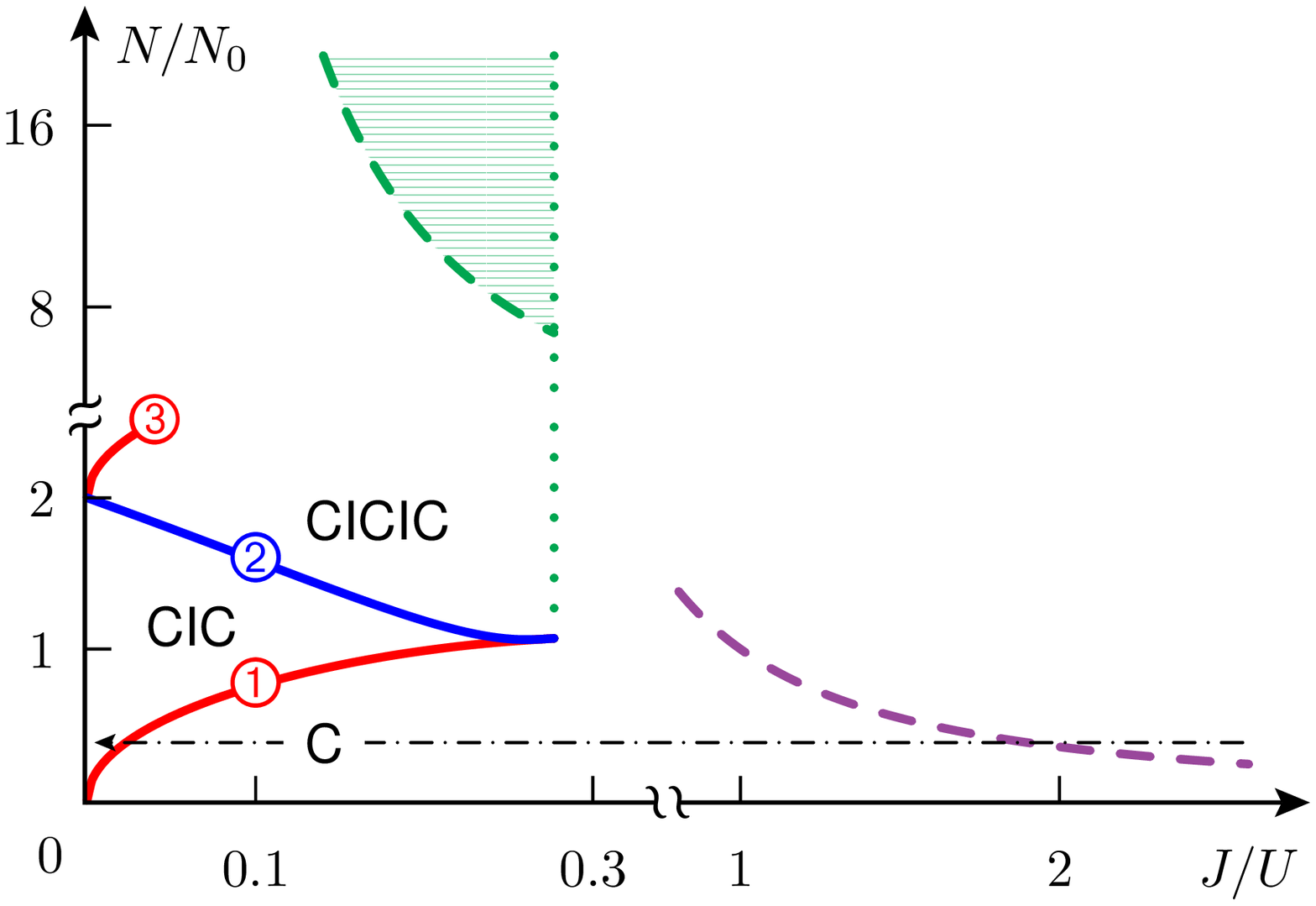}\caption{\label{phases}
 States of the trapped atomic cloud in the presence of optical lattice.
 Compressible (C) and incompressible (I) phases
 coexist in distinct states, such as CIC, etc.
 The solid line serves as boundary between these states.
 Its first two segments (denoted by \ding{192} and \ding{193})
 bound the state CIC, which consists of an incompressible
 region in the middle of the trap and two compressible 
 ones at the sides, see Fig.~\ref{motttrapfig}.
 The incompressible region nucleates at line \ding{192} 
 upon crossing over from state C by decreasing $J/U$ or increasing $N/N_0$.
 Similarly, traversing line \ding{193} upwards 
 leads to a compressible
 region form in the trap center; hence the combination CICIC.
 The small-$J$ asymptotes of lines \ding{192} and \ding{193}
 are given in Eqs.~(\ref{tgm}) and (\ref{NN0blueline}),
 respectively.
 The crossover line (dashed) at
 $N/N_0\lesssim 1$ is defined by Eq.~(\ref{wtos}) and separates the
 regimes of weak and strong interaction in the compressible state.
 At $N/N_0\gtrsim 1$, a mixed state
 with compressible and incompressible phases forms to the left of
 the vertical line (dotted). The compressible phases
 represent a substantial fraction of the cloud in the shaded
 region; its left boundary is given by Eq.~(\ref{wtos}). To the
 left of that line, the cloud is mostly distributed between incompressible
 states. We are interested in the evolution of the dynamic structure
 factor upon decreasing $J/U$ at small $N/N_0$, dash-dotted line.}
\end{figure}

\begin{figure}[t]
\includegraphics[width=0.95\columnwidth]{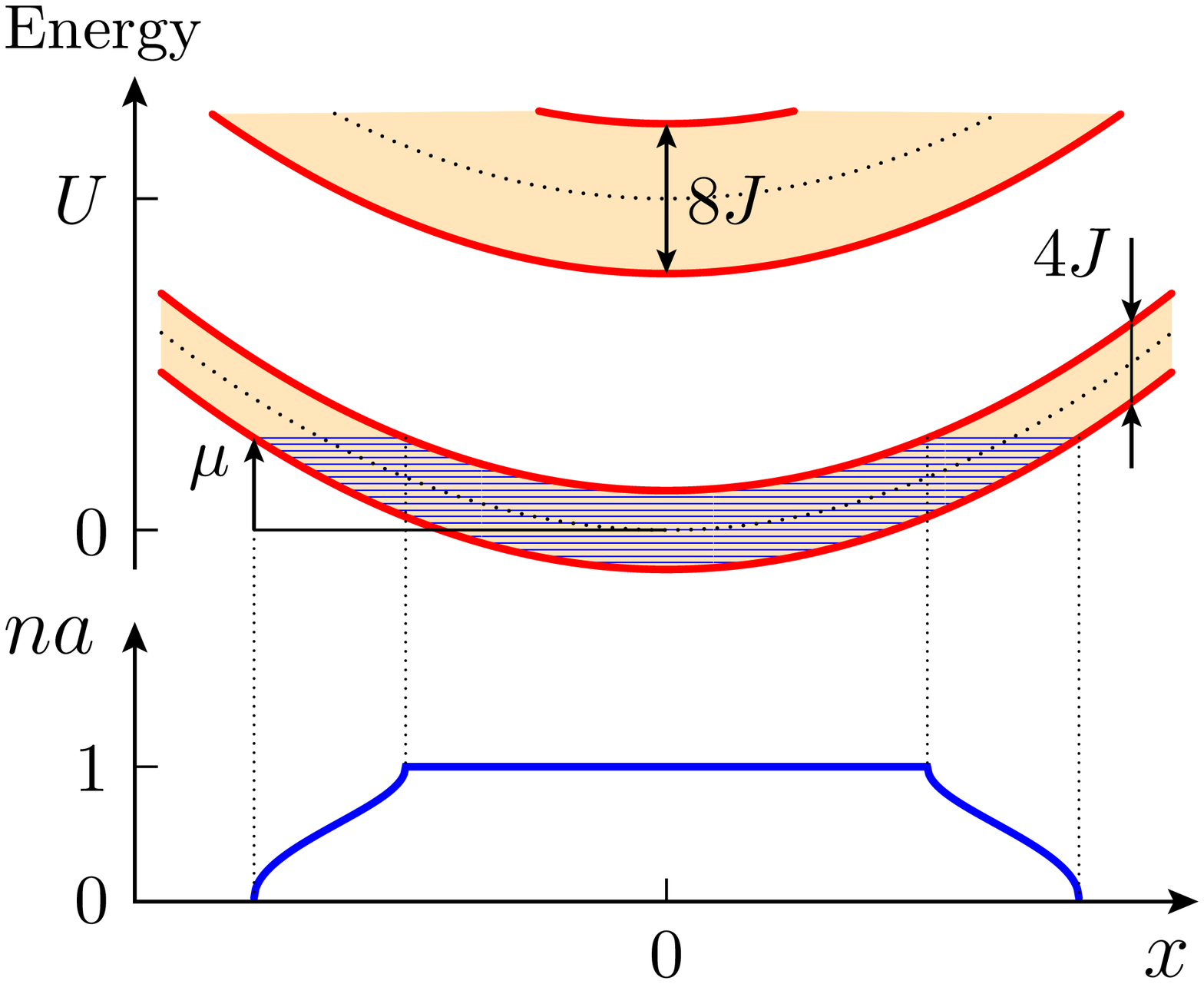}\caption{\label{motttrapfig}
Density profile and energy diagram of an atomic cloud in the Mott-insulator
regime confined in a trap.  Upper panel: Energy diagram showing two lowest Mott
bands curved due to the trap potential.  The hashed area shows the phase space
occupied by atoms below the level of the chemical potential $\mu$.  To leading
order of perturbation theory the bands are separated (center to center) by
energy $U$ and have widths of $4J$ and $8J$ for the lower and upper bands,
respectively.  The state of the liquid depicted here corresponds to the state
CIC of Fig.~\ref{phases}.  Lower panel: Density profile $n(x)$ of the atomic
cloud in the trap.  The incompressible region with $na=1$ occupies the center
of the trap for $2J<\mu<U-4J$ at $J/U\ll 1$ and it borders with two
compressible regions on the sides.  The compressible regions have $n(x)$ given
in Eq.~(\ref{nofxarcsin}).
}
\end{figure}

Let us consider first $N/N_0\ll 1$ and see the effect of decreasing $J/U$ on
the state of atomic cloud in the trap. At a sufficiently large $J/U$, the
dimensionless interaction parameter, $U/Jna$, is small throughout the trap
(here $n$ is the position-dependent atomic density). Using the weak-interaction
limit for the $n$-dependence of the chemical potential, $\mu=Una$, and the
Thomas-Fermi approximation, we find for the density $n_0$ at the center of the
trap,
\begin{equation}
n_0a= \left(\frac{3}{4}\frac{N}{N_0}\right)^{2/3}.
\label{bgas}
\end{equation}
For reasons which will become clear later we refer to this limit as the
Bogoliubov limit.  The crossover from the weak- to strong-interaction limit
occurs at the periphery of the cloud if $J$ is large, and moves towards the
center with the reduction of $J/U$. At $N\ll N_0$, the entire cloud enters the
strong-interaction limit before the Mott phase nucleates.  The
strong-interaction limit, $U/Jna\gg 1$, is described by the so-called
Tonks-Girardeau gas, for which the analog of Eq.~(\ref{bgas}) reads
\begin{equation}
n_0a= \frac{2^{1/2}}{\pi}\left(\frac{N}{N_0}\right)^{1/2}
\left(\frac{U}{J}\right)^{1/4},\; n_0a\ll 1.
\label{fgasn0a}
\end{equation}
The crossover between the Bogoliubov and Tonks-Girardeau limits occurs at
$U/Jn_0a\sim 1$.  Using the estimate for $n_0$ (either Eq.~(\ref{bgas}) or
Eq.~(\ref{fgasn0a})), we find for the crossover line
\begin{equation}
\frac{N}{N_0}=\left(\frac{U}{J}\right)^{3/2}.
\label{wtos}
\end{equation}
(Here we dispensed with the numerical factors present in Eqs.(\ref{bgas}) and
(\ref{fgasn0a}), as Eq.~(\ref{wtos}) defines a crossover rather than a
transition.)  The crossover line is shown in Fig.~\ref{phases} (see dashed line
at $N/N_0\lesssim 1$).  Further decrease of $J$ (moving along the dash-dotted
line in Fig.~\ref{phases}) eventually allows one to reach the boundary between
the Tonks-Girardeau gas and the Mott phase. The nucleation of the Mott phase
occurs in the center of the trap, once the condition
$N=(8/\pi)\sqrt{J/\epsilon_0}$ is reached;  in dimensionless variables, the
latter condition reads
\begin{equation}
\frac{N}{N_0}=\frac{8}{\pi}\left(\frac{J}{U}\right)^{1/2}.
\label{tgm}
\end{equation}
The corresponding boundary is shown in in Fig.~\ref{phases} by the solid line
(segment \ding{192}).  Under the condition $N/N_0\ll 1$, the Mott phase
occurs~\cite{footnote1} at $J/U\ll 1$.  The advantage of a small number of
particles in the trap, $N\ll N_0$, is that there is at most {\sl one} domain of
the trap occupied by the Mott phase, see Fig.~\ref{motttrapfig}.

For a large number of atoms, $N\gg N_0$, the composition of the atomic
cloud in the trap may become more complicated. Upon the reduction of
the bandwidth, the first incompressible phase
appears~\cite{Kuehner98,Rapsch99} at $J/U\lesssim 0.28$,
see dotted line in Fig.~\ref{phases}.
Further
reduction of the band width brings an increasing number of Mott phases
corresponding to various integer values $j$ of the filling, $n(l)a=j$.
Incompressible phases nucleate around multiple points in the trap,
where the condition $J/U\sim 1/j$ is satisfied~\cite{monien}. Once the
band width is reduced to satisfy the condition $J/U\sim (n_0a)^{-1}$,
a major part of the cloud belongs to the incompressible phases. The
value of $n_0a$ is estimated by Eq.~(\ref{bgas}) both in the limits of
weak ($J/U\gg 1$) and strong ($J/U\ll (n_0a)^{-1}$) interaction, where
the equation of state for incompressible phases is applicable. Using
Eq.~(\ref{bgas}), we may re-write the condition $J/U=(n_0a)^{-1}$
separating ``mostly compressible'' state from the ``mostly
incompressible'' one in the form of Eq.~(\ref{wtos}).

We summarize the important for us features of the phase diagram for
the trapped cloud subject to the optical lattice potential in
Fig.~\ref{phases}.
Reducing the parameter $J/U$ at $N/N_0\ll 1$ allows one to investigate
the gradual increase of correlations due to interaction in the liquid phase
as well as 
the formation of a single Mott phase domain in contact with Bose liquid 
(see Fig.~\ref{motttrapfig}). 
In the next section, we review the
manifestation of such an evolution of the state of the cloud in the
dynamic structure factor.

\subsection{Average dynamic structure factor in the absence of optical lattice}
\label{AvDyStFa}

In a homogeneous 1D system with contact interaction $g\delta (x_i-x_j)$ between
bosons (Lieb-Liniger model~\cite{LiebLiniger}), the dynamic response depends on
the dimensionless interaction constant $\gamma=mg/n$. In this section, we
analyze the density-averaged dynamic structure factor for a gas in a trap.

The dynamic structure factor at zero temperature is defined as
$S(q,\omega)={\rm Im}\langle\rho(x,t)\rho(0,0)\rangle_{q,\omega}$, where $\rho
(x,t)$ is the density fluctuation operator. In the Lieb-Liniger
model~\cite{LiebLiniger} the dynamic structure factor has a power-law
divergence~\cite{Khodas,Imambekov1} at the so-called~\cite{LiebLiniger}
Lieb-$1$ mode, $\omega=\varepsilon_+(q)$, and vanishes at the threshold
determined by the spectrum of the Lieb-$2$ mode: $S(q,\omega)= 0$ at 
$\omega < \varepsilon_-(q)$. 

The full form of $S(q,\omega)$ can be found analytically in the limit
of infinitely-strong contact repulsion between the bosons ($\gamma\gg
1$, limit of the Tonks-Girardeau gas):
\begin{eqnarray}
S(q,\omega)&=&\frac{m}{q}
\theta\left(\omega-\varepsilon_-(q)\right)
\theta\left(\varepsilon_+(q)-\omega\right),\nonumber\\
\varepsilon_{\pm}(q)&=&
\left|\frac{\pi nq}{m}\pm \frac{q^2}{2m}\right|.
\label{Sqwthetas}
\end{eqnarray}
If $\gamma\ll 1$, then the dispersion of Lieb-$1$ mode approaches the
Bogoliubov spectrum, 
\begin{equation}
\varepsilon_+(q)=\sqrt{\left(vq\right)^2+
\left(\frac{q^2}{2m}\right)^2},
\label{eqBogoSpectrum}
\end{equation}
with the collective velocity $v=\sqrt{gn/m}$. As one may expect,
$S(q,\omega)$ approaches a $\delta$-function,
\begin{equation}
S(q,\omega)= \frac{\pi nq^2}{m\varepsilon_+(q)}
\delta\left(\omega-\varepsilon_+(q)\right),
\label{Sqwdelta}
\end{equation}
in this limit.~\cite{Khodas} The Lieb-2 mode in the $\gamma\ll 1$
limit corresponds to the grey solitons spectrum.~\cite{Kulish} 
The spectral weight near $\varepsilon_-(q)$
is, however, extremely small~\cite{photosoliton} at any $q$.

At arbitrary $\gamma$, the general form of
$S(q,\omega)$ is~\cite{Khodas,Imambekov1}
\begin{equation}
S(q,\omega)\sim\frac{m}{q}
\left|\frac{\delta\epsilon}{\omega-\varepsilon_+}\right|^{\mu_1}
\left[\theta(\varepsilon_+-\omega)+\nu_1\theta(\omega-\varepsilon_+)\right]
\label{Sqwpowerlaw}
\end{equation}
with the exponent $\mu_1$ and asymmetry factor $\nu_1$ depending on
momentum $q$ and density $n$.

\begin{figure}[t]
\includegraphics[width=0.95\columnwidth]{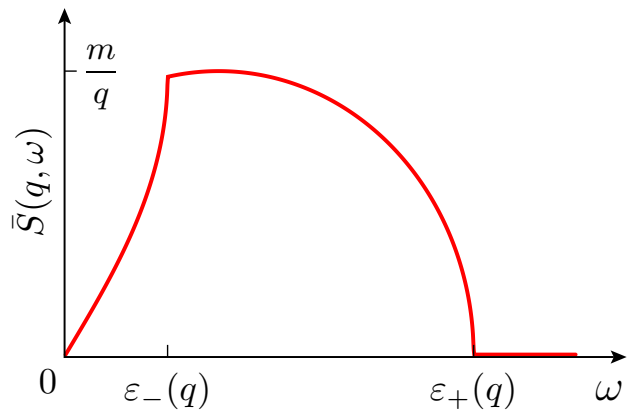}\caption{\label{sqwtgfig}
Dynamic response $\bar{S}(q,\omega)$ as a function of $\omega$ for an atomic
cloud in the Tonks-Girardeau limit, see Sec.~\ref{TGlimit}.  The
non-analyticities are at $\omega=\varepsilon_\pm(q)$, with $\varepsilon_\pm(q)$
given in Eq.~(\ref{Sqwthetas}).  The plot is made using Eq.~(\ref{bSmqsqrt})
for the choice of parameters $q=1.2 \pi n_0$.  In the presence of optical
lattice, one replaces $m\to m^*$.
}
\end{figure}

All the above forms of $S(q,\omega)$ were derived for a
spatially-homogeneous Bose liquid. How one may compare the theory for
a homogeneous system with measurements performed in the presence of a
trap?  We consider here the most favorable case of a smooth
longitudinal potential confining the 1D bosons, whose confinement
energy $\omega_0$ is negligibly small (not resolved on measurement
scale).  Due to the interaction between bosons, the system length $2L$
is usually much larger than the quantum length scale $\lambda$ of the
confinement, allowing for independent (unaffected by interference)
probing of different parts of the system.  Provided the system is
probed at a sufficiently large momentum, $q\gg 1/\lambda$, one may
regard portions of length $\lambda\sim 1/\sqrt{m\omega_0}$ as
homogeneous and responding independent of each other.  Then the
dynamic structure factor can be approximated by the density-averaged
one,
\begin{equation}
\bar{S}(q,\omega)=\frac{1}{2L}\int_{-L}^{L} dx 
S\left(q,\omega; n(x)\right).
\label{averdef}
\end{equation}
The confinement affects the dynamic response through the variation of the
density profile $n(x)$. The averaged structure factor $\bar{S}(q,\omega)$ may
differ considerably from $S(q,\omega)$ of a homogeneous system.  We note that
the accuracy of the density-averaged approximation applied to a smooth trapping
potential was checked in the limit $\gamma\to\infty$ and yielded excellent
results~\cite{VignoloMinguzziTosi} (error less than $5\%$).  Constraints on the
applicability of Eq.~(\ref{averdef}) may arise in some cases due to quantum
corrections to the dynamic response originating from momentum uncertainty on
the scale $\lambda^{-1}$.  We defer the analysis of this ``finite-size'' effect
to Appendix~\ref{AppFiniteSizeEffects}.

\begin{figure}[t]
\includegraphics[width=0.95\columnwidth]{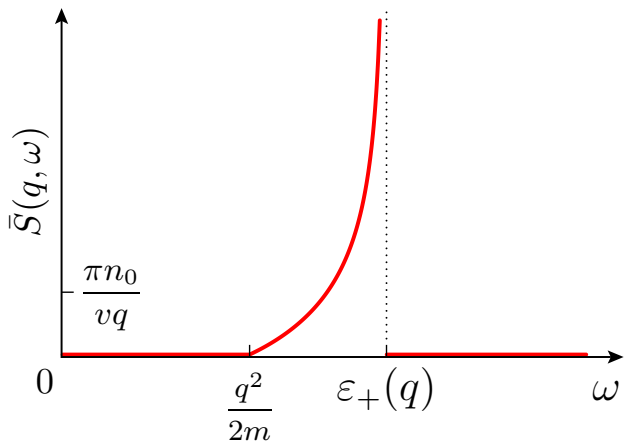}\caption{\label{sqwbogofig}
Dynamic response $\bar{S}(q,\omega)$ as a function of $\omega$ for an atomic
cloud in the Bogoliubov limit, see Sec.~\ref{Bogolimit}.  A square-root
singularity occurs at $\omega=\varepsilon_+(q)$, where $\varepsilon_+(q)$ is
given by the Bogoliubov spectrum in Eq.~(\ref{eqBogoSpectrum}).  The plot is
made using Eq.~(\ref{bSn02mv2q}) for the choice of parameters $q^2/2m=mv^2$.
In the presence of optical lattice, one replaces $m\to m^*$ and $g\to Ua$.
}
\end{figure}

For definiteness, we assume here a parabolic confinement, see Eq.~(\ref{conf}).
However, our main conclusions require only the existence of a smooth maximum in
the density profile, while the specific form of the $n(x)$ dependence is not
important.  For the parabolic confinement, averaging of Eq.~(\ref{Sqwthetas})
over the proper density profile yields~\cite{VignoloMinguzziTosi}
\begin{eqnarray}
\bar{S}(q,\omega)&=&\frac{m^2}{\pi n_0q^2}\left[
\sqrt{\left(\varepsilon_+(q)-\omega\right)
\left(\varepsilon_-(q)+\omega\right)}
\right. \nonumber\\
&&-\left.
\theta\left(\varepsilon_-(q)-\omega\right)
\sqrt{\left(\varepsilon_+(q)+\omega\right)
\left(\varepsilon_-(q)-\omega\right)}
\right]\nonumber\\
&&\times
\theta(\varepsilon_+(q)-\omega),
\quad\quad\quad\quad\quad\quad
\gamma\gg 1,
\label{bSmqsqrt}
\end{eqnarray}
where for brevity we included only interval of wavevectors $q\leq 2\pi n_0$
(see Sec.~\ref{TGlimit} for larger $q$).  The $\omega$-dependence of
$\bar{S}(q,\omega)$ in Eq.~(\ref{bSmqsqrt}) for a fixed value of $q$ is shown
in Fig.~\ref{sqwtgfig}.

Similarly, averaging of Eq.~(\ref{Sqwdelta}) yields
\begin{eqnarray}
\bar{S}(q,\omega)&=&\frac{\pi n_0}{mv^3q}
\frac{\omega^2-\left(q^2/2m\right)^2}{\sqrt{\varepsilon_+^2(q)-\omega^2}}
\theta(\varepsilon_+(q)-\omega)
\nonumber\\
&&
\times 
\theta(\omega-q^2/2m),
\quad\quad\quad\quad
\gamma\ll 1.
\label{bSn02mv2q}
\end{eqnarray}
The dependence of $\bar{S}(q,\omega)$ in Eq.~(\ref{bSn02mv2q}) on $\omega$ is
shown in Fig.~\ref{sqwbogofig}.  The parameters $n$ and $v$ in the dispersion
relation $\varepsilon_+(q)$, entering Eqs.~(\ref{bSmqsqrt})
and~(\ref{bSn02mv2q}), correspond to the maximal density, $n=n_0$.  While the
full form of the averaged structure factor substantially uses
Eqs.~(\ref{Sqwthetas}) and (\ref{Sqwdelta}) and assumes the parabolic
confinement, the behavior of $\bar{S}(q,\omega)$ near the upper edge,
$\omega\to\varepsilon_+(q)$, is universal and relies only on the form of the
corresponding singularity in $S(q,\omega)$ before the averaging, and on the
existence of quadratic maximum in the function $n(x)$.  Tuning of interaction
from weak to strong may be performed, for example, by decreasing the number of
atoms in the trap, which reduces $n_0$.  The response should then cross over
from the one with divergence at the upper edge $\varepsilon_+(q)$, see
Eq.~(\ref{bSn02mv2q}) and Fig.~(\ref{sqwbogofig}), to the one with a
non-analytic point at $\omega=\varepsilon_+(q)$ and maximum below
$\varepsilon_+(q)$, see Eq.~(\ref{bSmqsqrt}) and Fig.~(\ref{sqwtgfig}).

\begin{figure}[t]
\includegraphics[width=0.95\columnwidth]{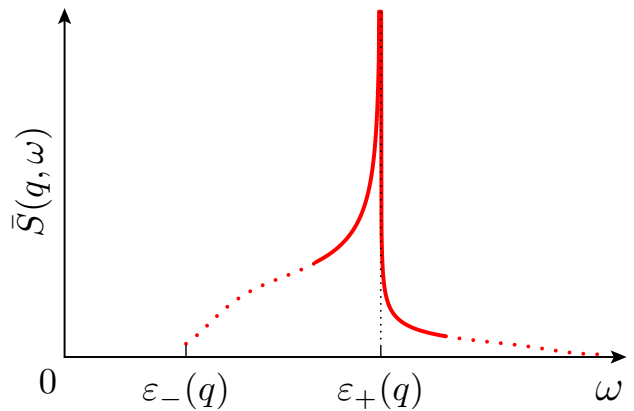}\caption{\label{sqwKfig}
Sketch showing the singular behavior of $\bar{S}(q,\omega)$ in
Eq.~(\ref{bSpowerlawABC}) at the Lieb-$1$ mode.  The choice of parameters
corresponds to the regime of ``large $q$ and arbitrary $\gamma$'' considered in
Sec.~\ref{limlargeqarbint}.  The dynamic response diverges at the Lieb-$1$ mode
$\varepsilon_+(q)$ (solid line) and is strongly suppressed away from
$\varepsilon_+(q)$ at distances exceeding $vq$ (dotted line).  The solid line
is plotted using Eq.~(\ref{longres1}) with the choice $\mu_1=0.8$; the dotted
line is hand-drawn and illustrates the qualitative behavior of
$\bar{S}(q,\omega)$ away from the Lieb-$1$ mode.
}
\end{figure}

At arbitrary interaction, $\bar{S}(q,\omega)$ demonstrates a
non-analytical behavior in the vicinity of $\omega=\varepsilon_+(q)$
of the general type
\begin{eqnarray}
\bar{S}(q,\omega)&\sim &
\left|\frac{\delta\epsilon}{\omega-\varepsilon_+(q)}\right|^{\mu_1-1/2}
\left[A\theta(\varepsilon_+(q)-\omega)
\right.\nonumber\\
&&\left.
+B\theta(\omega-\varepsilon_+(q))\right]
+C,
\label{bSpowerlawABC}
\end{eqnarray}
with a $q$-dependent exponent $\mu_1$ evaluated at maximum density
$n=n_0$; constants $A$, $B$, and $C$  are given in Section~\ref{ArbInt}.
We illustrate the singular behavior of $\bar{S}(q,\omega)$ 
at $\mu_1>1/2$ and $\nu_1=1$ in Fig.~\ref{sqwKfig}.
The divergence in $\bar{S}(q,\omega)$ disappears at $\mu_1<1/2$,
but the point $\omega=\varepsilon_+(q)$ remains non-analytic.

\subsection{The dynamic structure factor in the presence of optical
  latice, $N/N_0\ll 1$}
\label{secDSFinpsenceoflattice}

In the above consideration, we assumed no optical lattice, so that the
dispersion relation of 1D bosons is quadratic.  The presence of an optical
lattice changes the dispersion relation of individual bosons; sufficiently
strong lattice potential results in narrow bands of allowed energies.
Equations (\ref{bSmqsqrt})-(\ref{bSpowerlawABC}) remain true even in that case,
provided $N\ll N_0$ and $J/U\gg(N/2N_0)^2$, see Eqs.~(\ref{wtos}) and
(\ref{tgm}), and assuming $qa\ll 1$. In the narrow-band limit, the
free-particle mass $m$ in Eqs.~(\ref{bSmqsqrt})-(\ref{bSpowerlawABC}) must be
replaced by $1/2Ja^2$, where $4J$ is the bandwidth of the lowest allowed energy
band.  The general form of the singular contribution,
Eq.~(\ref{bSpowerlawABC}), is applicable also in the crossover regime specified
by Eq.~(\ref{wtos}).

The condition $J/U\gg(N/2N_0)^2$ ensures small occupation of the lattice sites,
$n_0a\ll 1$, which allows one the use of parabolic dispersion relation for
bosons. A reduction of the bandwidth eventually leads to occupation $n_0a\geq
1/2$ in the center of the trap, see Eq.~(\ref{fgasn0a}). Under this condition,
the inflection points in the full dispersion relation for a narrow-band
spectrum become important, leading to specific singularities in the dynamic
structure factor. These singularities, associated with symmetric configurations
of excited particles and holes, where studied for the homogeneous case in the
context of spin chains,~\cite{Niemeijer,Mueller,Pereira} and we will not dwell
on them here.

\begin{figure}[t]
\includegraphics[width=0.95\columnwidth]{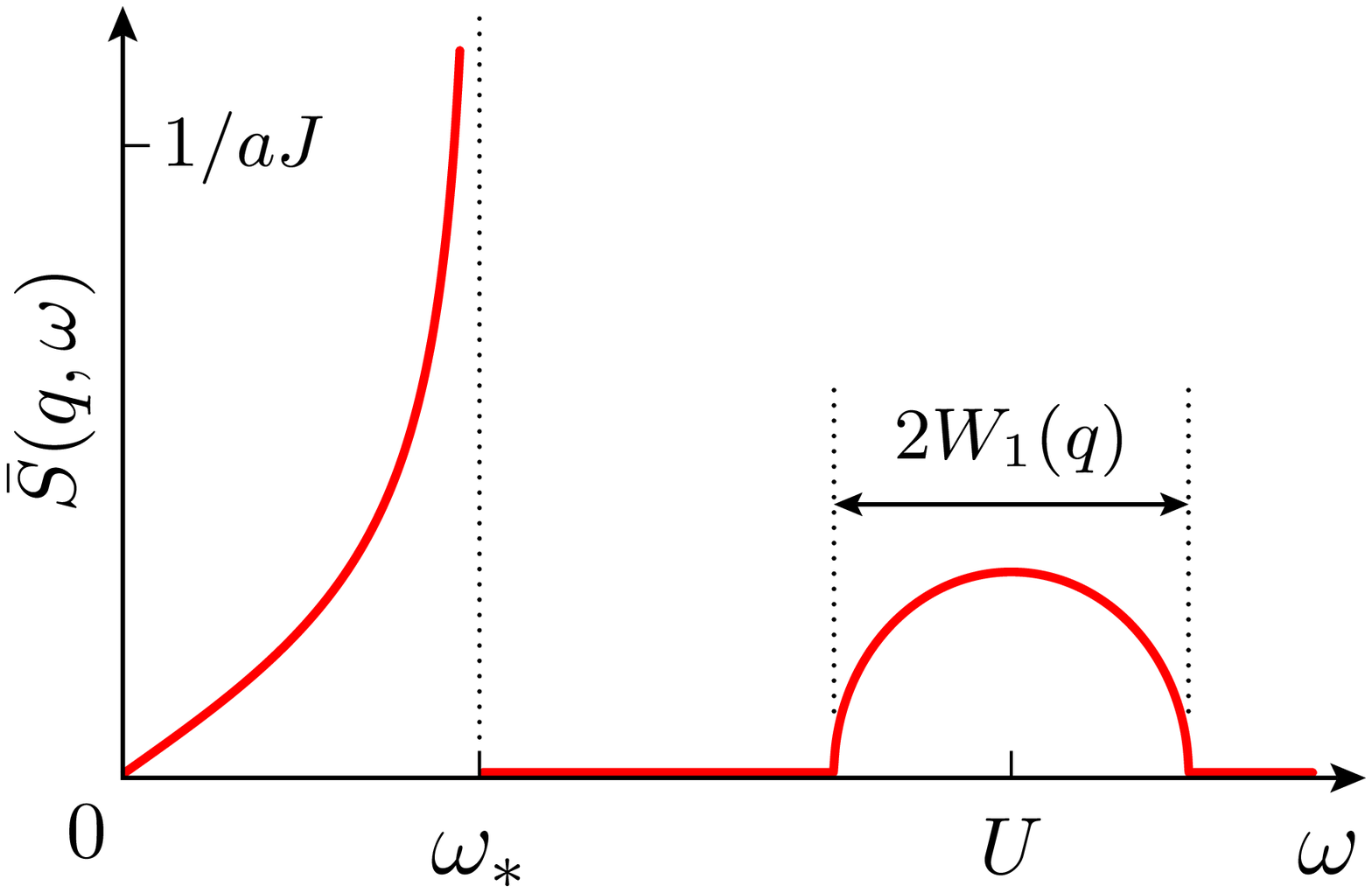}\caption{\label{sqwmottfig}
Dynamic response $\bar{S}=S_{\rm inc}+2\bar{S}_{\rm comp}$ as a function of
$\omega$ for an atomic cloud in the Mott regime, featuring a square root
singularity at frequency $\omega_*=4J\sin(qa/2)$ and a semi-circular peak at
$\omega=U$.  The plot is made using Eqs.~(\ref{bS2SbS}),
(\ref{Sincsemicircle}), and~(\ref{Scom}) with the parameters values: $q=\pi/a$,
$U=10J$, and $\epsilon_0N^2=10J$.  
}
\end{figure}

Further decrease of the bandwidth results in the formation of an incompressible
Mott phase in the center of the trap at $n_0a=1$.  The Mott phase turns, on
both sides, into compressible regions.  In these compressible caps, the
concentration $n$ drops from $n=1/a$ to zero continuously, see
Fig.~\ref{motttrapfig}.  The dynamic response is, therefore, equal to the sum
of contributions of the Mott phase (incompressible) and two caps
(compressible),
\begin{equation}
\bar{S}(q,\omega)=S_{\rm inc}(q,\omega)+ 
2\bar{S}_{\rm com}(q,\omega).
\label{bS2SbS}
\end{equation}
Here, the response is averaged over the trap with an appropriate density
profile, see Fig.~\ref{motttrapfig}, which accounts for the presence of the
compressible and incompressible parts.  At $J/U\ll (N/2N_0)^2$, the Mott phase
extends almost over the entire cloud, with the caps constituting a small part
of the system.  To the leading order in $J/U\ll 1$, the Mott phase contribution
is given by
\begin{equation}
S_{\rm inc}(q,\omega)=\frac{64J^2 \sin^2(qa/2)}{aU^2W_1(q)}
\sqrt{1-\left(\frac{\omega -U}{W_1(q)}\right)^2},
\label{Sincsemicircle}
\end{equation}
where $W_1(q)=2J\sqrt{1+8\cos^2\left({qa}/{2}\right)}$ and we assumed one atom
per site ($N\ll N_0$), see Section~\ref{SecPeakShape} for details.  Thus, at
the leading order of $J/U$, the Mott phase produces a semicircular peak of
width $2W_1(q)$, centered at $\omega=U$.  The contribution of the liquid caps
at frequencies $\omega\approx U$ can be safely neglected since it is small by a
factor $\Delta N/N\ll 1$, where $\Delta N=4J/\epsilon_0N$ is the number of
atoms in the caps.  At much smaller frequencies, $\omega\sim J$, the dynamic
response is solely due to the caps.  The contribution of one cap to
Eq.~(\ref{bS2SbS}) is
\begin{equation}
\bar{S}_{\rm com}(q,\omega)=
\frac{2\omega}{a\epsilon_0N^2}
\frac{\theta(\omega)\theta\left(4J\sin(qa/2)-\omega\right)}
{\sqrt{\left[4J\sin(qa/2)\right]^2-\omega^2}},
\label{Scom}
\end{equation}
see Section~\ref{SecCompRegion}. Here, the divergent contribution at the
threshold frequency comes from parts of the trap with occupation $na\approx
1/2$ and has the origin in the particle-hole
symmetry~\cite{Niemeijer,Mueller,Pereira} mentioned above.  The total dynamic
response in the Mott regime is shown in Fig.~\ref{sqwmottfig} and is plotted
using Eqs.~(\ref{bS2SbS}), (\ref{Sincsemicircle}), and~(\ref{Scom}).

The dynamic response $S(q,\omega)$ in the Mott regime ($J\ll U$) is peculiar,
featuring a hierarchy of weights at different frequencies.  The smallest
frequency scale is $\omega\sim J$, where the dynamic response is independent of
interaction at large $U$.  The other frequencies are integer multiples of $U$,
$\omega\approx jU$, where the dynamic response vanishes with increasing $U$ as
$\propto\left(J/U\right)^{2j}$, for $j=1,2,\dots$, as long as the applicability
of the Bose-Hubbard model holds.  Ignoring for simplicity the fine structure of
the peaks (each peak width is proportional to $J$), we represent the dynamic
structure factor as
\begin{equation}
\tilde{S}(q,\omega)=\frac{1}{a}\sum_{j=0}^{\infty}
\alpha_p\sin^{\nu(j)}(qa/2)
\left(\frac{J}{U}\right)^{2j}\delta\left(\omega-\Omega_j(q)\right),
\label{manymotts}
\end{equation}
where $\nu(0)=1$ and $\nu(j>0)=2$,
$\Omega_0= 4J\sin(qa/2)$ and $\Omega_{j>0}=jU$, and 
$\alpha_j$ are numerical coefficients.

A possible way to tune between different states of a 1D Bose system is provided
by control of the optical lattice potential.  Typically~\cite{Bloch} the
crossover between the limits of weak and strong interaction occurs for a
sufficiently strong optical lattice potential, allowing one the use of the
tight-binding approximation for the single-particle spectrum and of the 1D
Bose-Hubbard model for the description of interaction. The increase of the
optical lattice potential leads chiefly to the decrease of the single-particle
band width $4J$, while the on-site repulsion $U$ changes little. Narrowing of
the band width $4J$ leads to the evolution of the dynamic structure factor.
This evolution allows for a clear interpretation at $N\ll N_0$, see
Sec.~\ref{SecPhases}. We will follow the changes in the dynamic response
accompanying the reduction of the $J/U$ ratio at small $N$, {\it e.g.}, along
the dash-dotted line in Fig.~\ref{phases}.

The evolution between the regimes of weak interaction, strong interaction in
the compressible phase, and the regime of coexistence of compressible and Mott
phases is best illustrated by the integral characteristics of the dynamic
response. For the weakly-interacting gas at $qa\ll 1$, we find
\begin{equation}
\int\frac{d\omega}{2\pi}\bar{S}(q,\omega)=
\frac{3^{1/3}\pi}{8\cdot 2^{1/6}}\left(\frac{N}{N_0}\right)^{1/3}
\left(\frac{J}{U}\right)^{1/2} q,
\end{equation}
while after the crossover to the Tonks-Girardeau limit one obtains
\begin{equation}
\int\frac{d\omega}{2\pi}\bar{S}(q,\omega)=
\frac{q}{2\pi}.
\label{weight-Fermi}
\end{equation}
The above two estimates match each other at the crossover line in
Eq.~(\ref{wtos}). Remarkaby, in the Tonks-Girardeu limit the integral
intensity of the dynamic response is independent of the parameters of the
system.

Upon further bandwidth reduction, the Mott phase is formed, and at
$J/U\ll (\pi N/8N_0)^2$, it occupies the most part of the trap, in which case
\begin{eqnarray}
\int \frac{d\omega}{2\pi} S_{\rm inc}(q,\omega)
   &=&
4\left(\frac{J}{U}\right)^2 a q^2,\;\quad qa\ll 1,\label{weight-inc}\\
\int \frac{d\omega}{2\pi} \;2\bar{S}_{\rm com}(q,\omega)
   &=&
\frac{4}{\pi}\left(\frac{N_0}{N}\right)^2\frac{J}{U} q.
\label{weight-com}
\end{eqnarray}
Note that Eqs.~(\ref{weight-Fermi}) and (\ref{weight-com}) match each other (up
to a numerical factor) at the transition line in Eq.~(\ref{tgm}).  Comparing
Eqs.~(\ref{Scom}) and (\ref{weight-com}) with Eqs.~(\ref{Sincsemicircle}) and
(\ref{weight-inc}), respectively, we make two interesting observations.  First,
at $J/U\ll (N/N_0)^2$ the responses of the compressible and incompressible
parts of the cloud occur at well-separated intervals of $\omega$.  Second, the
total weight of the response of the Mott phase is smaller by the factor
$\pi(N/N_0)^2(J/U)(aq)$ than the weight provided by the compressible phases,
although the latter occupy only a small fraction of the atomic cloud.

The decrease of the integral intensity in the Mott phase does match the
experimental observations, and does not contradict sum rules, see
Sec.~\ref{SecfSumRules}. Indeed the major part of the spectral weight in the
presence of the lattice is shifted to higher frequencies associated with the
transitions from the lowest to higher single-particle bands.

\section{Dynamic Structure Factor of the Homogeneous Mott Phase}
\label{DSFofHMPh}

In this section we consider the Mott phase of the Bose-Hubbard mode and
evaluate the dynamic response to leading order of perturbation theory in
$J/U\ll 1$.  The trap potential $V(x)$ causes no inhomogeneity of the Mott
phase and affects only the length over which the Mott phase extends.
Considering here a macroscopically large Mott phase (extending over many
lattice sites), we omit the trap potential $V(x)$ from Eq.~(\ref{BH}) and
separate the Hamiltonian into a main part ($H_0$) and a perturbation ($H_1$) as
follows
\begin{eqnarray}
\label{HBHsecond}
H_{\rm BH}&=&H_0+H_1,\\
\label{H0}
H_0&=&\frac{U}{2}\sum_l\left(b_l^\dagger b_l-\eta\right)^2,\\
\label{H1}
H_1&=&-J\sum_l\left(b_{l+1}^\dagger b_l+b_l^\dagger b_{l+1}\right).
\end{eqnarray}
Here, $\eta$ controls the occupancy of a lattice site
and is related to the chemical potential $\mu$
in a linear fashion,
$\eta=1/2+\mu/U$.

Upon raising the chemical potential $\mu$ the lattice begins to be occupied at
$\mu=-2J$, and up to $\mu=2J\left[1-2J/U+{\cal O}\left((J/U)^2\right)\right]$,
the state of the lattice is compressible, turning as next into a Mott-insulator
state with site occupancy one.  The Mott-insulator state turns to a
compressible state at $\mu=U-4J\left[1+{\cal O}\left(J/U\right)\right]$ and
emerges again at $\mu=U+4J\left[1+{\cal O}\left(J/U\right)\right]$ with site
occupancy two.  We shall be interested, in this section,  in any Mott-insulator
state, which amounts to choosing values of $\eta$ away from half-integers by a
value $\sim J/U$.  More precisely, to leading order in $J/U$, the state of the
cloud is a Mott insulator if 
\begin{equation}
\left|\mu-\nu U\right|>2(\nu +1)J,
\label{MottPhaseCriterium}
\end{equation}
for all $\nu = 0,1,2\dots$. The integer $\nu$ for which l.h.s. in
Eq.~(\ref{MottPhaseCriterium}) takes the smallest value, gives the site
occupation number,
\begin{equation}
p=\nu+\frac{1}{2}+\frac{1}{2}\,{\rm sgn}\,(\mu - \nu U).
\label{occupnump}
\end{equation}

The dynamic structure factor for the Bose-Hubbard model
is defined as follows
\begin{equation}
S(q,\omega)=\frac{1}{a}\sum_l\int_{-\infty}^{\infty}d\tau
e^{i\omega\tau-iqal}
\left\langle
\rho_l(\tau)\rho_{0}
\right\rangle,
\label{defSqwHBM}
\end{equation}
where $\rho_l=b_l^\dagger b_l-\langle b_l^\dagger b_l\rangle$ is the deviation
of the occupation number of site $l$ from its average value and $\rho_l(\tau)$
denotes the Heisenberg representation of $\rho_l$.

Before going into the details of a rigorous derivation of the peak structure of
$S(q,\omega)$ at $\omega\approx U$, we outline first a simpler derivation which
gives the correct spectral weight of the peak. The derivation of the peak
shape is deferred to Section~\ref{SecPeakShape}.

\subsection{Spectral weight at $\omega\approx U$}
\label{SecSpecralWeight}

For a Mott-insulator state, the ground state in perturbation theory reads
\begin{equation}
|\tilde\Psi_0\rangle=
\left(1-\frac{H_1}{U}\right)|\Psi_0\rangle+{\cal O}(H_1^2),
\label{eqstn2}
\end{equation}
where $|\Psi_0\rangle$ stands for the
unperturbed ground state
\begin{equation}
|\Psi_0\rangle=\prod_{l}\frac{(b_l^\dagger)^p}{\sqrt{p!}}\left|0\right\rangle,
\label{eqstn1}
\end{equation}
with $p$ being the Mott state occupation number.  Note that the form of
Eq.~(\ref{eqstn2}) is owing to the fact that $H_1$ acting on $|\Psi_0\rangle$
creates only one kind of excitations.  The excitation energy is approximately
$U$ and each excitation is represented by a particle-hole pair with hard-core
interaction between the particle and the hole.

Since here we are interested only in the spectral weight of $S(q,\omega)$, it
suffices to write down a complete set of states spanning the space of all
excitations with energy $\omega\approx U$ in the lowest order of perturbation
theory. Orthogonality with respect to Eq.~(\ref{eqstn2}) implies
\begin{equation}
|\tilde\Psi_{l_1l_2}\rangle=
|\Psi_{l_1l_2}\rangle
+|\Psi_0\rangle\frac{\langle \Psi_0|H_1|\Psi_{l_1l_2}\rangle}{U}+{\cal O}(H_1^2),
\label{eqstatl1l2tilde}
\end{equation}
where $|\Psi_{l_1l_2}\rangle$ stands for an unperturbed excited state with a
particle at site $l_1$ and a hole at site $l_2\neq l_1$,
\begin{equation}
|\Psi_{l_1l_2}\rangle=
\frac{1}{\sqrt{p(p+1)}}
b_{l_1}^\dagger b_{l_2}
|\Psi_0\rangle, \quad\quad (l_1\neq l_2).
\label{eqstatl1l2one}
\end{equation}

Evaluating first the matrix elements
\begin{eqnarray}
\langle\tilde\Psi_0|
b_l^\dagger b_l
|\tilde\Psi_{l_1l_2}\rangle&=&\frac{J\sqrt{p(p+1)}}{U}
\left(\delta_{l_1, l_2+1}+\delta_{l_2, l_1+1}\right)\nonumber\\
&&\times
\left(\delta_{l, l_1}-\delta_{l, l_2}\right),
\end{eqnarray}
we obtain for the dynamic structure factor in 
Eq.~(\ref{defSqwHBM}) the following result
\begin{equation}
S(q,\omega)=
\frac{16\pi 
p(p+1)}{a}\left(\frac{J}{U}\right)^2\sin^2\left(qa/2\right)\delta\left(\omega-U\right).
\label{Sqwppdeltasw}
\end{equation}
Note that the peak weight increases with $p$ like $p^2$ and is suppressed at
small $J$ by the factor $(J/U)^2$.  By generalizing the above derivation, it
becomes clear that there exist also peaks in $S(q,\omega)$ at frequencies equal
to integer multiples of $U$, i.e. $\omega\approx j U$, however the weight of
those peaks is suppressed as $(J/U)^{2j}$.

The $\delta$-function in Eq.~(\ref{eqstn1}) comes from the fact that we
neglected the dispersion relation of the excitations, assuming that they all
have energy $\omega=U$.  As a matter of fact, accounting for the dispersion
relation of the excitations results only in a broadening of the $\delta$-peak,
without changing its spectral weight, see next Section.

\subsection{Semi-circular peak at $\omega\approx U$}
\label{SecPeakShape}
The states in Eq.~(\ref{eqstatl1l2tilde}) span the space of all excitations at
energy $\omega\approx U$.  However, they are not eigenstates of the Hamiltonian
(\ref{HBHsecond}), not even at zeroth order of $J/U$.  Below, we build up the
eigenstates of the Hamiltonian (\ref{HBHsecond}) by a suitable linear
combination of the states (\ref{eqstatl1l2tilde}).  In order to do so, we first
consider the two-body problem that represents the excitation.  We solve this
problem by going to the center-of-mass and relative motion coordinates.  Then,
having proper states at the zeroth order of $J/U$, i.e. analogs of
Eq.~(\ref{eqstatl1l2one}), we repeat the steps in Sec.~\ref{SecSpecralWeight}
to obtain the dynamic structure factor.

To simplify our notations in this section, we set the lattice spacing to unity
($a=1$) and agree to denote any coordinate dependence by a subscript and
momentum dependence by a superscript.  Our starting point are the states in
Eq.~(\ref{eqstatl1l2one}) and the matrix elements of the Hamiltonian
(\ref{HBHsecond}) in the space spanned by these states.  The interaction part
$H_0$ is already diagonal
\begin{equation}
\langle\Psi_{p_1p_2}|
H_0
|\Psi_{l_1l_2}\rangle=
\delta_{p_1l_1}\delta_{p_2l_2}
\left[U+\frac{U}{2}\sum_l(p-\eta)^2\right],
\end{equation}
and moreover, it is proportional to the unity matrix, and thus, is invariant
under unitary transformations.  We leave $H_0$ aside for the time being and
focus on the tunneling part
\begin{eqnarray}
\langle\Psi_{p_1p_2}|
H_1
|\Psi_{l_1l_2}\rangle&=&
-J\left[
(p+1)\delta_{p_2l_2}\left(\delta_{p_1,l_1+1}+\delta_{p_1,l_1-1}\right)
\right.\nonumber\\
&&
\left.+
p\delta_{p_1l_1}\left(\delta_{p_2, l_2+1}+\delta_{p_2,l_2-1}\right)
\right].
\label{meVppll}
\end{eqnarray}
We decouple the motion of the center of mass~\cite{noteCMM}
of the excitation by the following transformation
\begin{equation}
|\Psi_{r}^{\varkappa}\rangle=
\frac{1}{\sqrt{2\pi}}\sum_{l\in I_r}
e^{i\varkappa l/2}
\left|\Psi_{\frac{l+r}{2},\frac{l-r}{2}}
\right\rangle,
\label{sttCMM}
\end{equation}
where $r=\pm 1,\pm 2, \pm 3,\dots$, the sum over $l$ runs over all integers of
the same parity as $r$ (hence the notation $l\in I_r$), and
$\varkappa\in[-\pi,\pi]$.  The states in Eq.~(\ref{sttCMM}) are normalized as
usually,
\begin{equation}
\langle\Psi_{r'}^{\varkappa'}
|\Psi_{r}^{\varkappa}\rangle=\delta_{r'r}\delta(\varkappa'-\varkappa).
\end{equation}
In the new basis, the center of mass motion decouples and the matrix becomes
diagonal in the momentum $\varkappa$,
\begin{eqnarray}
\langle\Psi_{r'}^{\varkappa'}|
H_1
|\Psi_{r}^{\varkappa}\rangle
&=&
{\cal H}_{r'r}(\varkappa)
\delta(\varkappa-\varkappa').
\end{eqnarray}
The resulting relative-motion problem reads
\begin{equation}
{\cal H}_{r'r}(\varkappa)=
-T_\varkappa\delta_{r',r-1}
-T_\varkappa^*\delta_{r',r+1},
\label{effeprobcalHrr}
\end{equation}
where $r',r=\pm 1,\pm 2,\pm 3,\dots$ and the effective tunneling
amplitude is given by
\begin{equation}
T_\varkappa=J\left(
pe^{-i\varkappa/2}
+(p+1)e^{i\varkappa/2}
\right).
\end{equation}
The momentum $\varkappa$ enters in Eq.~(\ref{effeprobcalHrr}) as a parameter.
Note that the range of values of $r',r$ in Eq.~(\ref{effeprobcalHrr}) does not
include zero. We can, therefore, split ${\cal H}_{r'r}$ into two semi-infinite
chains, ${\cal H}_{r'r}={\cal H}_{r'r}^{+}+{\cal H}_{r'r}^{-}$, where ${\cal
H}_{r'r}^{\pm}$ is given by Eq.~(\ref{effeprobcalHrr}) with solely
positive/negative values of $r',r$.  Each of the semi-infinite chains can be
solved easily by fictitiously including zero into the domain of values of
$r',r$, but requiring that the wave function vanishes at site $r=0$. The
solution then reads
\begin{eqnarray}
\left|\Psi^{k\varkappa}_{\pm}\right\rangle=
\sqrt{\frac{2}{\pi}}
\sum_{r=1}^{\infty}e^{\mp ir\phi_\varkappa}\sin(kr)
\left|\Psi_{\pm r}^\varkappa\right\rangle,
\label{stkkppa}
\end{eqnarray}
were $k\in (0,\pi)$ and the sign index $\pm$
stands for the right/left chain.
The phase $\phi_\varkappa$ originates from
$T_\varkappa=|T_\varkappa|e^{i\phi_\varkappa}$
with
\begin{equation}
|T_\varkappa|=J\sqrt{1+4p(p+1)\cos^2(\varkappa/2)}.
\label{Tvarkappa}
\end{equation}
The states in Eq.~(\ref{stkkppa}) are normalized as follows
\begin{equation}
\left\langle\Psi^{k'\varkappa'}_{s'}|
\Psi^{k\varkappa}_{s}\right\rangle=\delta_{s's}\delta(k'-k)\delta(\varkappa'-\varkappa).
\end{equation}
Taking into account the part $H_0$,
we obtain that the particle-hole excitation
has the dispersion relation
\begin{equation}
\varepsilon_\pm(k,\varkappa)=U-2|T_\varkappa|\cos(k),
\label{disprelepskkappa}
\end{equation}
with $|T_\varkappa|$ given in Eq.~(\ref{Tvarkappa}).

The next step would be to correct the states
$\left|\Psi^{k\varkappa}_{\pm}\right\rangle$
in perturbation theory to the first
order of $J/U$, and thus, obtain the analog
of the states in Eq.~(\ref{eqstatl1l2tilde}).
This can be done straightforwardly, and
for reference, the result reads
\begin{equation}
|\tilde\Psi_\pm^{k\varkappa}\rangle=
|\Psi_\pm^{k\varkappa}\rangle-
\frac{2J\sqrt{p(p+1)}}{U}\sin(k)\delta(\varkappa)|\Psi_0\rangle.
\end{equation}
It is, however, not necessary to repeat the same steps 
as in Sec.~\ref{SecSpecralWeight}---remarkably, all those steps
can be comprised in a single formula and one
can use the bare states $\left|\Psi^{k\varkappa}_{\pm}\right\rangle$
to evaluate matrix elements.
We use the following expression,
which turns out to be a considerable 
shortcut in the calculation,
\begin{eqnarray}
\langle\tilde\Psi_0|
b_l^\dagger b_l
|\tilde\Psi^{k\varkappa}_{\pm}\rangle=
\frac{i}{U}\langle \Psi_0|{\cal J}_l\left|\Psi^{k\varkappa}_{\pm}\right\rangle,
\label{shortcut}
\end{eqnarray}
where ${\cal J}_l$ is the current operator
${\cal J}_l=i[ H,\; b_l^\dagger b_l ]$
which reads explicitly
\begin{equation}
{\cal J}_l=iJ\left(
b_{l}^\dagger b_{l+1}-
b_{l+1}^\dagger b_l\right)+
iJ\left(
b_{l}^\dagger b_{l-1}-
b_{l-1}^\dagger b_l\right).
\end{equation}
Equation (\ref{shortcut}) holds for the considered
model to first order in  $J/U$.

Using Eq.~(\ref{shortcut}) or repeating the 
steps in Sec.~\ref{SecSpecralWeight}, we finally obtain
\begin{eqnarray}
\langle\tilde\Psi_0|
b_l^\dagger b_l 
|\tilde\Psi_\pm^{k\varkappa}\rangle&=&
\pm\frac{2J\sqrt{p(p+1)}}{i\pi U}e^{\mp i\phi_\varkappa+i\varkappa l}\nonumber\\
&&\times
\sin(k)\sin(\varkappa/2).
\label{matelkappaks}
\end{eqnarray}
With the help of the matrix elements in Eq.~(\ref{matelkappaks})
and the dispersion relation in Eq.~(\ref{disprelepskkappa}),
the dynamic structure factor in Eq.~(\ref{defSqwHBM}) evaluates to
\begin{eqnarray}
S(q,\omega)&=&\frac{2p(p+1)}{a}\left(\frac{4J}{U}\right)^2\sin^2(qa/2)\nonumber\\
&&\times
\frac{1}{W_p(q)}\sqrt{1-\left(\frac{\omega -U}{W_p(q)}\right)^2}.
\label{Sincsemicirclepp}
\end{eqnarray}
Here, we have reintroduced the lattice spacing $a$
and denoted the width of the peak by $2W_p(q)$, where
\begin{equation}
W_p(q)=2J\sqrt{1+4p(p+1)\cos^2(qa/2)}.
\label{Wofq}
\end{equation}
This completes the derivation of Eq.~(\ref{Sincsemicircle}), which is readily
obtained from Eq.~(\ref{Sincsemicirclepp}) by specifying the value $p=1$ for
the Mott lobe occupancy.  Note also that the spectral weight of $S(q,\omega)$
in Eq.~(\ref{Sincsemicirclepp}) is equal to the spectral weight of
$S(q,\omega)$ in Eq.~(\ref{Sqwppdeltasw}), as expected from our discussion in
Sec.~\ref{SecSpecralWeight}.

Finally, we note that the same problem was studied in Ref.~\onlinecite{Rey} in
the limit $p\gg 1$ using a combination of perturbation theory and periodic
boundary condition on the wave function. The result of Ref.~\onlinecite{Rey}
for $S(q,\omega)$ represents a series of closely spaced $\delta$-spikes located
in the neighborhood of $\omega=U$.  These spikes become increasingly denser and
can be replaced by a continuous curve in the limit of large period of the
boundary condition.  In order to compare our result against the result of
Ref.~\onlinecite{Rey}, we replace the sum in Eq.~(9) of Ref.~\onlinecite{Rey}
by an integral (an extra factor of $1/2$ is needed to account for the sparse
summation in the sum).  After integration, we recover a semicircular peak in
$S(q,\omega)$ centered at $\omega=U$. The width of the peak agrees with our
result for width in the limit $p\cos\left(qa/2\right)\gg 1$; the prefactor of
$S(q,\omega)$ agrees as well, provided differences in the definition of
$S(q,\omega)$ are taken into account.

\section{Averaged Dynamic Structure Factor}

\label{SecAverDynStrFactor}
\noindent
In this section, we study the effect of inhomogeneity of the atomic cloud on
the dynamic structure factor $S(q,\omega)$.  The inhomogeneity is caused by the
trap potential $V(x)$, see Eq.~(\ref{conf}).  The characteristic length over
which the atomic density $n(x)$ varies along the trap is usually much larger
than the quantum length scale $\lambda$ of the longitudinal trap confinement.
For such ``soft'' confinements, we employ the density averaging in
Eq.~(\ref{averdef}).  We first find $n(x)$ in the regimes of interest, using
the Thomas-Fermi approximation, and then evaluate the dynamic structure factor
averaged over the trap inhomogeneity.

\subsection{Atomic density profile in a 1D trap}

\label{SecAtmDens1Dtr}
\noindent
Here, we consider a 1D Bose gas subjected to the potentials of a harmonic trap
and an optical lattice.  We assume that the lattice potential is strong enough
to justify our use of the Bose-Hubbard model in Eq.~(\ref{BH}).  As a limiting
case of the Bose-Hubbard model (continuum limit), we recover the model of 1D
Bose gas subjected to only a harmonic trap potential, see end of this
subsection.

For a soft trap confinement, the density profile $n(x)$ is commonly found using
the Thomas-Fermi approximation.  The trap potential $V(x)$ is treated
classically and taken into account through the local electro-chemical balance
\begin{equation}
\frac{\partial {\cal E} }{\partial n}+
V(x)=\mu,
\label{elchembalance}
\end{equation}
where ${\cal E}$ is the ground state energy (per unit volume) of a homogeneous
system of density $n$. In calculating ${\cal E}$, energy should be measured
with respect to the chemical potential.  The profile $n(x)$ is found by solving
Eq.~(\ref{elchembalance}) for $n$.  This approximation yields $n(x)$ accurately
for the most part of the cloud, with exceptions being the phase boundaries
(points at which $na=0,1,2\dots$), where the approximation works only
qualitatively.  For the purpose of calculating the dynamic response of the
whole cloud Eq.~(\ref{elchembalance}) is sufficient, because only the gross
features of $n(x)$ matter.

For weak interaction $U/Jn_0a\ll 1$ (Bogoliubov limit), due to the bosonic
nature of the problem, the resulting density profile is due to the balance
between mean-field interaction energy and external harmonic trapping. It is
convenient to measure energies from the bottom of the band, introducing
$\bar{\mu}=\mu+2J$. The density profile is then given by
\begin{equation}
n(x)=\frac{\bar{\mu}-V(x)}{Ua}=n_0\left[1-\left(\frac{x}{L}\right)^2\right],
\label{nofxboseUtna}
\end{equation}
where $n_0=\bar{\mu}/Ua$ is the atomic density at the trap center, and $2L$ is
the cloud length, with $L=a\sqrt{\bar{\mu}/\epsilon_0}$.  For the Bose-Hubbard
model (\ref{BH}), Eq.~(\ref{nofxboseUtna}) is obtained from
Eq.~(\ref{elchembalance}) in the limit $J/U\gg 1$, using perturbation theory in
the interaction.  At smaller bandwidth, $J/U\lesssim 1$,
Eq.~(\ref{nofxboseUtna}) continues to describe the gross features of $n(x)$, as
long as $U/Jn_0a\ll 1$, see discussion in Sec.~\ref{SecPhases}.  Equation of
state (relation between $\bar{\mu}$ and $N$) is obtained from
Eq.~(\ref{nofxboseUtna}) by integration, i.e. imposing the normalization
condition on the density profile, $N=\int_{-L}^{L} n(x) dx$, yielding
\begin{equation}
N/N_0=\frac{4}{3}\left(\bar{\mu}/U\right)^{3/2},
\label{eqofstatebosons}
\end{equation}
where $N_0$ is defined in Eq.~(\ref{N0}). Upon substitution of
$\bar{\mu}=n_0aU$ in Eq.~(\ref{eqofstatebosons}), we recover Eq.~(\ref{bgas}).

In the opposite limiting case $U/Jn_0a\gg 1$, the Thomas-Fermi approximation is
equivalent to the original Thomas-Fermi approximation for fermions, except for
the fact that several layers of atoms can be added on top of each other at the
price of raising the chemical potential by about $U$ for each layer.  For
example, the second layer of atoms begins to form in the center of the trap
when the following condition is satisfied (retaining leading order in $J/U\ll
1$)
\begin{equation}
U-4J=\epsilon_0\left(N/2\right)^2,
\label{U4tepsN2}
\end{equation}
where $4J$ represents half of the width of the second Mott band (see
Fig.~\ref{motttrapfig}) and $N/2$ represents $L/a$ at the leading order of
$J/U$.  In terms of the total number of atoms $N$ and in the limit $J\to 0$,
the second layer starts forming with increasing $N$ at $N=2N_0$.  In this
section, we will consider not more than a single layer of atoms in the trap,
which can be rephrased, with the help of Eq.~(\ref{U4tepsN2}), as follows
\begin{equation}
N/N_0 < 2\sqrt{1-4J/U}\approx 2\left(1-2J/U\right).
\label{NN0blueline}
\end{equation}
Precisely this linear dependence of the maximal $N/N_0$ on $J/U$ is displayed
in Fig.~\ref{phases} at small $J/U$ (see line \ding{193}).  Note that
Eq.~(\ref{NN0blueline}) is accurate only to first order of $J/U$, because we
used leading order expressions for the Mott bandwidth and atomic cloud length
in Eq.~(\ref{U4tepsN2}).

At $J/U\ll 1$  (Tonks-Girardeau limit),  the solution can be found by mapping
the bosons onto noninteracting fermions, where the repulsive on-site
interaction can be accounted for by the Pauli exclusion principle.  The atomic
density $n(x)$ in the compressible region is then found from
\begin{equation}
2J\left[1-\cos\left(\pi a n(x)\right)\right]+V(x)=\bar{\mu},
\label{eqn2t2cosenbalance}
\end{equation}
where we have used the tight-binding expression for the chemical potential of a
homogeneous system of non-interacting fermions.  In the incompressible region,
Eq.~(\ref{eqn2t2cosenbalance}) does not hold and $n(x)$ is trivially given by
$n(x)=1/a$. From Eq.~(\ref{eqn2t2cosenbalance}), we find the well-known
expression for the atomic density
\begin{equation}
n(x)=\frac{2}{\pi a}\arcsin\sqrt{\frac{\bar{\mu}-V(x)}{4J}}.
\label{nofxarcsin}
\end{equation}
As before, the equation of state is found by integrating $n(x)$ over $x$.  At
$0<\bar{\mu}< 4J$, the atomic cloud consists of a single compressible phase
(state C in Fig.~\ref{phases}). From Eq.~(\ref{nofxarcsin}), we obtain
\begin{equation}
N=\frac{4}{\pi}\sqrt{\frac{4J-\bar{\mu}}{\epsilon_0}}
\left[
E\left(\frac{\bar{\mu}}{\bar{\mu}-4J}\right)-
K\left(\frac{\bar{\mu}}{\bar{\mu}-4J}\right)
\right],
\label{NEllipticEK}
\end{equation}
where $K(m)$ and $E(m)$ are the complete elliptic integrals of the first and
second kind, respectively.  At $4J<\bar{\mu}<U-4J$, a Mott phase is present in
the middle of the cloud, see Fig.~\ref{motttrapfig}.  Adding the contributions
of the Mott phase and two compressible caps to the total number of atoms, we
obtain
\begin{equation}
N=\frac{4}{\pi}\sqrt{\frac{\bar{\mu}}{\epsilon_0}}
E\left(\frac{4J}{\bar{\mu}}\right), 
\quad\quad \bar{\mu}\geq 4J.
\label{NEllipticE}
\end{equation}
At the lower limit of applicability of Eq.~(\ref{NEllipticE}), $\bar{\mu}=4J$,
a Mott phase nucleates in the center of the trap; with $E(1)=1$ in
Eq.~(\ref{NEllipticE}), one obtains Eq.~(\ref{tgm}) for $N$.

In the low-filling regime  $\bar{\mu}\ll 4J$ we recover the continuum limit.
An arbitrary interaction strength can be accounted for within the Lieb-Liniger
model\cite{LiebLiniger} for bosons with effective mass defined by optical
lattice.  In this limit, all atoms reside at the bottom of the band, $na\ll 1$,
and one can approximate the tight-binding dispersion relation as follows
\begin{equation}
\varepsilon(k)=2J\left(1-\cos\left(ka\right)\right)\approx \frac{k^2}{2m^*},
\end{equation}
where $m^*=1/2Ja^2$ is the effective mass in the continuum model.  In the weak
interaction limit, $n(x)$ is given by Eq.~(\ref{nofxboseUtna}).  At strong
interactions, $U/Jn_0a\gg 1$, one obtains from the fermionic representation of
the model that
\begin{equation}
n(x)=n_0\sqrt{1-\left(\frac{x}{L}\right)^2},
\label{nofxsqrt}
\end{equation}
where $n_0=(1/\pi a)\sqrt{\bar{\mu}/J}$ and $L=a\sqrt{\bar{\mu}/\epsilon_0}$.
In both cases, $n(x)$ is quadratic in $x$ around the trap center---a feature
that also persists through the crossover regime.\cite{Dunjko} The equation of
state is obtained by integrating $n(x)$. At $U/Jn_0a\gg 1$, integration of
Eq.~(\ref{nofxsqrt}) yields
\begin{equation}
N=\frac{\bar{\mu}}{2\sqrt{J\epsilon_0}}.
\label{Ncontferm}
\end{equation}
Upon substitution $\bar{\mu}=J\left(\pi n_0 a\right)^2$
into Eq.~(\ref{Ncontferm}), we recover Eq.~(\ref{fgasn0a}).
We note that Eqs.~(\ref{nofxsqrt}) and (\ref{Ncontferm}) are
applicable at arbitrary ratios $J/U$, as long
as $U/Jn_0a\gg 1$, which means 
$\bar{\mu}/J\ll \min\left\{1, (U/J)^2\right\}$.

Having discussed the continuum limit of lattice bosons, it is easy to adapt
Eqs.~(\ref{nofxboseUtna}), (\ref{eqofstatebosons}), (\ref{nofxsqrt}), and
(\ref{Ncontferm}) to the case of confined bosons without an optical lattice.
It is achieved by the replacement: $m^*\to m$ and $Ua\to g$.

\subsection{Dynamic structure factor in the continuum limit, Bogoliubov regime}
\label{Bogolimit}

\noindent
For sufficiently small values of $U$, the interaction strength is small,
$U/Jn_0a\ll 1$, and the chemical potential lies within the lower parabolic part
of the spectrum in the Bose-Hubbard model, $\bar{\mu}\ll 4J$.  Considering also
small transition frequencies $\omega\ll 4J$, we may take the continuum limit,
$a\sim 1/\sqrt{J}\sim 1/U \sim \sqrt{\epsilon_0} \to 0$, in the Bose-Hubbard
model and match it with the model for the Bogoliubov sound in 1D, with the only
difference being the presence of the trap potential $V(x)$.  The Hamiltonian
(\ref{BH}) can thus be represented by
\begin{eqnarray}
H&=&\int dx
\phi^\dagger(x)
\left[-\frac{1}{2m^*}\frac{\partial^2}{\partial x^2}
+V(x)-\bar{\mu}\right]
\phi(x)\nonumber\\
&&+
\frac{Ua}{2}\int dx
\phi^\dagger(x)\phi^\dagger(x)
\phi(x)\phi(x),
\label{HcontBG}
\end{eqnarray}
where $1/m^*=2Ja^2$ and $\phi(x)$ is a boson field operator normalized as
$[\phi(x),\phi^\dagger(x')]=\delta(x-x')$.  The model in Eq.~(\ref{HcontBG})
coincides with the model we discussed in Sec.~\ref{AvDyStFa}, up to obvious
notational differences: $m\leftrightarrow m^*$ and $g\leftrightarrow Ua$.

In the homogeneous case ($V(x)=0$), the dynamic structure factor at small $U$
is sharply peaked at the frequency of the Bogoliubov spectrum, see
Eqs.~(\ref{eqBogoSpectrum}) and (\ref{Sqwdelta}).  In order to understand the
limit in which Eq.~(\ref{Sqwdelta}) is valid, one should imagine taking $U\to
0$ while maintaining $\bar{\mu}$ (i.e. $Una$) constant.  Then,
Eq.~(\ref{Sqwdelta}) is exact.  In practice, Eq.~(\ref{Sqwdelta}) is a fairly
good approximation to the dynamic structure factor at finite $U$ for small
$U/Jna\ll 1$, since a weak interaction only broadens the $\delta$-function in
Eq.~(\ref{Sqwdelta}) to a power-law singularity, see Sec.~\ref{ArbInt}.

Next, we average Eq.~(\ref{Sqwdelta}) with the help of Eq.~(\ref{averdef}),
using the Thomas-Fermi density $n(x)$ in Eq.~(\ref{nofxboseUtna}).  We express
the result through the parameters at the trap center and, to avoid doubling
notations, we redefine now $v$ as follows, $v=\sqrt{n_0aU/m^*}$.  The atomic
density at the trap center $n_0$ is given in Eq.~(\ref{bgas}).  We obtain
Eq.~(\ref{bSn02mv2q}), with $m\to m^*$.

As mentioned above, the Bogoliubov limit is the limit of small $U$.
The requirement  $U/Jn_0a\ll 1$ can be rewritten using the crossover boundary
in Eq.~(\ref{wtos}) as
\begin{equation}
N\gg N_0\left(\frac{U}{J}\right)^{3/2},
\label{NggN032Utbogo}
\end{equation}
whereas the requirement $\bar{\mu}\ll 4J$ 
can be rewritten using Eq.~(\ref{eqofstatebosons}) as
\begin{equation}
N\ll N_0\left(\frac{J}{U}\right)^{3/2}.
\end{equation}
The two requirements can be met only if $J/U\gg 1$.  Furthermore, in order to
justify the use of Eq.~(\ref{averdef}), we have to require that the size of the
cloud be much larger than the quantum length scale associated with the motion
of a band atom in the trap potential.  This imposes a condition on $N$ which
becomes more rigid and replaces Eq.~(\ref{NggN032Utbogo}) at $U\lesssim
t^{3/4}\epsilon_0^{1/4}$,
\begin{equation}
N\gg \frac{1}{\sqrt{N_0}}\left(\frac{J}{U}\right)^{3/4}.
\end{equation}

We summarize here our result: the singular dependence of $S(q,\omega)$ on
$\omega$, present at weak interactions in the 1D model, is smeared out towards
lower frequencies.  Thus, the $\delta$-function divergence present in
Eq.~(\ref{Sqwdelta}) is weakened to a square root divergence as shown in
Eq.~(\ref{bSn02mv2q}).

\subsection{Dynamic structure factor in the continuum limit, Tonks-Girardeau regime}
\label{TGlimit}

\noindent
Next we consider the limit of strong interaction, $U/Jn_0a\gg 1$, and, for
simplicity, restrict our consideration to small energies, $\bar{\mu}, \omega
\ll 4J$.  Following the same steps as in Sec.~\ref{Bogolimit}, we arrive at the
Hamiltonian (\ref{HcontBG}) in the continuum limit.  In the limit of strong
interaction, the Hamiltonian (\ref{HcontBG}) maps onto a free fermion model
(Tonks-Girardeau gas)
\begin{equation}
H=\int dx
\psi^\dagger(x)
\left[-\frac{1}{2m^*}\frac{\partial^2}{\partial x^2}
+V(x)-\bar{\mu}\right]
\psi(x),\quad
\label{HcontTG}
\end{equation}
where $\psi(x)$ is a fermion field operator normalized as follows
$\{\psi(x),\psi^\dagger(x')\}=\delta(x-x')$.  Note that the interaction
parameter $U$ is not present in Eq.~(\ref{HcontTG}), because the effect of the
repulsive interaction is accounted for by the fermionic nature of $\psi(x)$.
The correspondence between $\psi(x)$ and the bosonic field $\phi(x)$ in
Eq.~(\ref{HcontBG}) involves a non-local phase factor.  However, the latter
cancels out in quantities which do not involve permutation of particles.  In
particular, the phase factor cancels out in the dynamic structure factor, since
for the density operator we have
$\phi^\dagger(x)\phi(x)=\psi^\dagger(x)\psi(x)$.


For a non-interacting model, the dynamic structure factor can be easily
calculated for an arbitrary $V(x)$.  However, since we are interested here in
the limit of weak trap confinement, we employ nevertheless the density
averaging in Eq.~(\ref{averdef}).  In the homogeneous case ($V(x)=0$), the
dynamic structure factor is given by Eq.~(\ref{Sqwthetas}), with $m\to m^*$ and
$\pi n\equiv k_F=\sqrt{2m^*\bar{\mu}}=(1/a)\sqrt{\bar{\mu}/J}$.  We note that
Eq.~(\ref{Sqwthetas}) is exact in the limit $U/Jna\to \infty$ and it is a
fairly accurate approximation at any finite but large interaction strength
$U/Jna\gg 1$, provided the considered momenta are small $q\ll k_F(U/Jna)$, see
Sec.~\ref{ArbInt}.

Averaging Eq.~(\ref{Sqwthetas}) with the help of Eq.~(\ref{averdef}) and using
$n(x)$ in Eq.~(\ref{nofxsqrt}), we find $\bar{S}(q,\omega)$ and express it
through the parameters at the trap center.  To avoid doubling notations, we
redefine now $k_F$ to refer to the trap center, $k_F=\pi n_0$.  The averaged
dynamic structure factor is then given by Eq.~(\ref{bSmqsqrt}) for 
$q\leq 2\pi n_0$. At larger momenta, $q> 2\pi n_0$, a similar expression reads
\begin{eqnarray}
\bar{S}(q,\omega)&=&\frac{{m^*}^2}{\pi n_0q^2}
\sqrt{
\left(\varepsilon_+(q)-\omega\right)
\left(\omega-\varepsilon_-(q)\right)
}\nonumber\\
&&\times
\theta(\varepsilon_+(q)-\omega)
\theta(\omega-\varepsilon_-(q)).
\label{Sqwtgroot1}
\end{eqnarray}

The Tonks-Girardeau limit is the limit of small densities.
The requirement $U/Jn_0a\gg 1$ can be rewritten 
using the crossover boundary in Eq.~(\ref{wtos}) as
\begin{equation}
N\ll N_0\left(\frac{U}{J}\right)^{3/2},
\end{equation}
whereas the requirement $\bar{\mu}\ll 4J$ 
(or equivalently $n_0a\ll 1$) can be rewritten 
using Eq.~(\ref{Ncontferm}) as
\begin{equation}
N\ll N_0\left(\frac{J}{U}\right)^{1/2}.
\label{NllN0toU1o2}
\end{equation}
This condition is compatible with the condition of having
no Mott phase in the trap, see Eq.~(\ref{tgm}).
Additionally, in order to justify our use of Eq.~(\ref{averdef})
in the Tonks-Girardeau limit we need to require that $N\gg 1$,
which guarantees that the length of the atomic cloud is
much larger than the trap quantum length.

Next we analyze Eq.~(\ref{bSmqsqrt})  
in the limit of small momenta, $q\ll k_F$.
In this limit the change due to the averaging is dramatic.
The spectral weight of $S(q,\omega)$, contained before
the averaging within $\varepsilon_-(q)<\omega<\varepsilon_+(q)$,
is nearly fully pushed out into the region $0<\omega<\varepsilon_-(q)$.
We compare the spectral weights in the two regions in what follows.
To leading order in $q/k_F\ll 1$, Eq.~(\ref{bSmqsqrt}) reads
\begin{equation}
\bar{S}(q,\omega)=
\frac{m^*}{q}
\sqrt{2-\frac{2m^*\omega}{k_F q}+\frac{q}{k_F}},
\label{bSsqrt21mW}
\end{equation}
for $\left|\omega- k_F q/m^*\right|\leq q^2/2m^* $,
and
\begin{eqnarray}
\bar{S}(q,\omega)&=&
\frac{2{m^*}^2\omega}{k_F^2q}
\left[1+\frac{m^*\omega}{k_F q}\right]^{-1/2}
\left[\sqrt{1-\frac{m^*\omega}{k_F q}+\frac{q}{2k_F}}
\right.  \nonumber\\
&&+\left.
\sqrt{1-\frac{m^*\omega}{k_F q}-\frac{q}{2k_F}}\right]^{-1},
\label{bSsqrt21mWlong}
\end{eqnarray}
for $0\leq \omega\leq k_F q/m^* - q^2/2m^* $.
The net spectral weight in Eq.~(\ref{bSsqrt21mW}) is given by
\begin{equation}
\int_{\varepsilon_-}^{\varepsilon_+}\frac{d\omega}{2\pi} S(q,\omega)=
\frac{q}{3\pi}\sqrt{\frac{2q}{k_F}}.
\label{TGSWeight1}
\end{equation}
To the same order in $q/k_F$, the net spectral weight in 
Eq.~(\ref{bSsqrt21mWlong}) reads
\begin{equation}
\int_{0}^{\varepsilon_-}\frac{d\omega}{2\pi} S(q,\omega)=
\frac{q}{2\pi}-
\frac{q}{3\pi}\sqrt{\frac{2q}{k_F}}.
\label{TGSWeight2}
\end{equation}
Comparing Eqs.~(\ref{TGSWeight1}) and~(\ref{TGSWeight2}) to each other,
we find that the spectral weight remaining in the region $\varepsilon_-<\omega<\varepsilon_+$
constitutes only a small part $\sim \sqrt{q/k_F}$ of the total weight $q/2\pi$.
Note that the total weight before averaging is also $q/2\pi$.
This behavior can be described on a simpler level as follows.

Let us ignore the structure of the (rectangular) peak in 
Eq.~(\ref{Sqwthetas}) and replace it by a $\delta$-function with the same weight,
\begin{equation}
S(q,\omega)=q\delta(\omega-k_Fq/m^*).
\label{TGSqwqdelta}
\end{equation}
This approximation can be rigorously justified in the limit $q\to 0$, 
because the peak width tends to zero faster ($\propto q^2$) 
than the frequency at which the peak is centered ($\propto q$).
Carrying out the averaging for $S(q,\omega)$ in Eq.~(\ref{TGSqwqdelta}),
we obtain
\begin{equation}
\bar{S}(q,\omega)=\frac{m^*}{k_F}\frac{\omega}{\sqrt{\left(k_Fq/m^*\right)^2-\omega^2}}.
\label{TGSqwqdeltaAVERAGE}
\end{equation}
Equation~(\ref{TGSqwqdeltaAVERAGE}) coincides with
Eq.~(\ref{bSsqrt21mWlong}) to leading order in $q/k_F$
in the whole range of $\omega$, except for the neighborhood
of $\omega=k_Fq/m^*$.
A more precise condition for the validity of Eq.~(\ref{TGSqwqdeltaAVERAGE})
reads $|\omega-k_Fq/m^*|\gg q^2/m^*$.
Equations~(\ref{TGSqwqdelta}) and~(\ref{TGSqwqdeltaAVERAGE})
explain the gross features of the redistribution of the spectral weight
present in Eqs.~(\ref{bSsqrt21mW}) and~(\ref{bSsqrt21mWlong}).
Note that the square-root singularity present in Eq.~(\ref{TGSqwqdeltaAVERAGE})
resembles the one obtained for the Bogoliubov limit in Eq.~(\ref{bSn02mv2q}).
Although the qualitative behaviors of Eqs.~(\ref{bSn02mv2q}) and~(\ref{TGSqwqdeltaAVERAGE})
are similar,
the Bogoliubov and Tonks-Girardeau limits can be distinguished from each other
at small $q$ by the $\omega$-dependence in the numerators of 
Eqs.~(\ref{bSn02mv2q}) and~(\ref{TGSqwqdeltaAVERAGE}).

We summarize here our results.
The step-like behavior of $S(q,\omega)$ at 
$\omega=\varepsilon_+(q)$ present in the homogeneous
system is smeared to a square-root dependence 
$\propto\sqrt{\varepsilon_+-\omega}$,
with $\varepsilon_+(q)$ evaluated at the density of the
trap center.
The step at
$\omega=\varepsilon_-(q)$ is smeared in the same way
($\sqrt{\omega-\varepsilon_-}$) for $q\geq 2k_F$.
For $q<2k_F$, 
the smearing leads to 
non-zero spectral weight in the region
$0<\omega<\varepsilon_-(q)$; the behavior of 
$\bar{S}(q,\omega)$ at $\omega=\varepsilon_-(q)$ remains non-analytic,
with a diverging derivative on the lower frequency side.

\subsection{Dynamic structure factor in the continuum limit, at arbitrary interaction strength}
\label{ArbInt}

\noindent
In the limits of weak and strong interaction, the dynamic structure factor
$\bar{S}(q,\omega)$ turns to zero if $\omega$ exceeds the Lieb-$1$ frequency
$\varepsilon_+(q;n_0)$, corresponding to the maximal density $n_0$ in the trap
(see Secs.~\ref{Bogolimit} and~\ref{TGlimit}).  Contrary to this limiting
behavior, at intermediate interaction strength, $\bar{S}(q,\omega)\neq 0$ at
$\omega> \varepsilon_+(q;n_0)$.  Such behavior of $\bar{S}(q,\omega)$ for
trapped atoms reflects the evolution of $S(q,\omega)$, with the interaction
strength, in a homogeneous system.

In the Bogoliubov limit, a homogeneous system is characterized
by a $\delta$-function response, see Eq.~(\ref{Sqwdelta}).
Deviations from that weak-interaction
limit leads to a replacement
of the $\delta$-function singularity by a power-law
divergence~\cite{Khodas,Imambekov1} with an exponent $\mu_1<1$,
see Eq.~(\ref{Sqwpowerlaw}).
The singularity remains nearly symmetric in 
$\omega-\varepsilon_+(q)$ and the exponent close
to one, if the interaction parameter,
\begin{equation}
\gamma=\left\{
\begin{array}{ll}
mg/n, & \quad \mbox{without optical lattice},\\
m^*Ua/n, & \quad \mbox{with optical lattice},
\end{array}
\right.
\label{gammadefgamma}
\end{equation}
is small, $\gamma\ll 1$.
Upon increase of $\gamma$,
an asymmetry develops in the peak structure
(the higher frequency part, $\omega>\varepsilon_+(q)$,
becomes suppressed), and $\mu_1$ is decreasing
towards zero.
At $\gamma\gg 1$ (Tonks-Girardeau limit),
the shape of $S(q,\omega)$ approaches
$\theta(\varepsilon_+ -\omega)$,
see Eq.~(\ref{Sqwthetas}).

In the following, we study how the 
singular behavior of $S(q,\omega)$ at Lieb-$1$ mode
changes due to averaging over the density
$n(x)$ of the atomic cloud.
Our basic assumption is that $n(x)$ is a smooth
function of $x$ and can be 
approximated around its maximum at the trap center
by a quadratic expansion
\begin{equation}
n(x)\approx n_0+\frac{1}{2}n''(0) x^2,
\label{nofxexpand}
\end{equation}
where $n_0=n(0)$.
We note that this assumption 
about $n(x)$ agrees 
with the numerical analysis of Ref.~\onlinecite{Dunjko}.

The behavior of $S(q,\omega)$ at Lieb-$1$ mode
is summarized in Eq.~(\ref{Sqwpowerlaw}), see also 
Refs.~\onlinecite{Khodas,Imambekov1}.
We find that the divergence in $S(q,\omega)$ at Lieb-$1$ mode 
is weakened due to averaging over the trap.
We average $S(q,\omega)$ in Eq.~(\ref{Sqwpowerlaw})
using $n(x)$ in Eq.~(\ref{nofxexpand})
and obtain $\bar{S}(q,\omega)$ in Eq.~(\ref{bSpowerlawABC})  
with the coefficients:
\begin{eqnarray}
A&=&\frac{2\pi\mu_1\Lambda}{\sin(2\pi\mu_1)}\left[\cos(\pi\mu_1)-\nu_1\right],\nonumber\\
B&=&-\frac{\pi\mu_1\nu_1\Lambda}{\cos(\pi\mu_1)},\nonumber\\
C&=&\frac{\nu_1\Lambda}{1-2\mu_1}.
\label{ABCcoefs}
\end{eqnarray}
Here, the common factor $\Lambda$ reads
\begin{equation}
\Lambda=\frac{cm}{qL}
\left[\frac{\delta\epsilon n_0}{2 \Delta\varepsilon |n''(0)|}\right]^{1/2},
\label{LambdaL}
\end{equation}
where
$c=\sqrt{\pi}/\Gamma(1+\mu_1)\Gamma(3/2-\mu_1)$
is a factor order unity ($1\leq c < 2.2$) and
$\Delta\varepsilon$ is a scale defined below Eq.~(\ref{bSintegral1}).
The nature of non-analytic behavior of $\bar{S}(q,\omega)$
is established by Eqs.~(\ref{bSpowerlawABC}), (\ref{ABCcoefs}) 
and~(\ref{LambdaL}) for arbitrary $\gamma$.
Next we provide a more detailed form of $\bar{S}(q,\omega)$,
which can be matched to the results of Sections~\ref{Bogolimit} and~\ref{TGlimit}.

\subsubsection{Limit of large $q$ and arbitrary $\gamma$}
\label{limlargeqarbint}

\noindent
Here, we consider $q\gg q_0$, with 
\begin{equation}
q_0=
\left\{
\begin{array}{ll}
m^*v,&\quad \gamma\lesssim 1,\\
\gamma\, k_F, &\quad \gamma\gtrsim 1.
\end{array}
\right.
\label{q0expl}
\end{equation}
In this limit, 
the dynamic structure factor close to Lieb-$1$ mode is given by~\cite{Khodas,Imambekov1}
(cf. Eq.~(\ref{Sqwdelta}))
\begin{equation}
S(q,\omega)\simeq \frac{Kq}{\varepsilon_+(q)}
\delta_{1-\mu_1}\left(\frac{\omega-\varepsilon_+(q)}{v q}\right),
\label{Sqwdeltaarbint}
\end{equation}
where
$K=\pi n/m^*v$ is the Luttinger liquid parameter, $\mu_1=1-1/2K$,
and
$\delta_\epsilon(x)=(\epsilon/2)\left|x\right|^{\epsilon-1}$. 
Note that $\lim_{\epsilon\to 0}\delta_\epsilon(x)=\delta(x)$.
The Luttinger liquid parameter $K$ is a function of $\gamma$ only
and, for the Lieb-Liniger model, it has the asymptotes~\cite{LiebLiniger}
\begin{equation}
K=\left\{
\begin{array}{ll}
\pi/\sqrt{\gamma}, &\quad \gamma\ll 1,\\
1+4/\gamma, &\quad \gamma\gg 1.
\end{array}
\right.
\end{equation}
The equality sign in Eq.~(\ref{Sqwdeltaarbint}) holds 
for the Bogoliubov limit $\gamma\to 0$, whereas
at finite $\gamma$ 
Eq.~(\ref{Sqwdeltaarbint}) gives
$S(q,\omega)$ by order of magnitude
in a finite range around Lieb-$1$ mode,
$\left|\omega-\varepsilon_+(q)\right|\lesssim vq$.
Note that at $\gamma\gg 1$, parameter $K\to 1$ and $v\to \pi n/m^*$.

Substituting Eq.~(\ref{Sqwdeltaarbint}) into Eq.~(\ref{averdef})
and using Eq.~(\ref{nofxexpand}) we arrive at
\begin{equation}
\bar{S}(q,\omega)\sim 
\frac{q}{4\varepsilon_+L}\sqrt{\frac{n_0vq}{2|n''(0)|\Delta\varepsilon}}
\int_0^{\infty}\frac{d\xi}{\sqrt{\xi}}\left|\frac{1}{W+\xi}\right|^{\mu_1},
\label{bSintegral1}
\end{equation}
where $\Delta\varepsilon=n_0(\partial\varepsilon_+/\partial n_0)$ and 
$W=(\omega-\varepsilon_+)/vq$.
In obtaining Eq.~(\ref{bSintegral1}), 
we linearized the $n$-dependence of the Lieb-$1$ mode,
\begin{equation}
\varepsilon_+(q, n)\approx \varepsilon_+(q, n_0)+ \frac{\partial \varepsilon_+(q, n_0)}{\partial n_0}(n-n_0),
\end{equation}
and retained $n$-dependence only in the position of the singularity in Eq.~(\ref{Sqwdeltaarbint}).
The upper limit of integration is taken to infinity, since
keeping it finite goes beyond the accuracy of these approximations.
Equation~(\ref{bSintegral1}) gives the leading divergence
(at $\mu>1/2$) in $\bar{S}(q,\omega)$ at $W\to 0$.
All parameters in Eq.~(\ref{bSintegral1}) refer to the trap center.

Performing the integration in Eq.~(\ref{bSintegral1}), we obtain
\begin{eqnarray}
\bar{S}(q,\omega)&\sim&\frac{q}{4\varepsilon_+L}
\sqrt{\frac{n_0vq}{2|n''(0)|\Delta\varepsilon}}
\left\{
\left|\frac{vq}{\omega-\varepsilon_+}\right|^{\mu_1-1/2}
\right.\nonumber\\
&&
\times
\frac{\sqrt{\pi}\Gamma(\mu_1-\frac{1}{2})}{\Gamma(\mu_1)}
\left[
\tan\left(\frac{\pi\mu_1}{2}\right)
\theta(\varepsilon_+-\omega)
\right.\nonumber\\
&&\left.\left.
+\theta(\omega-\varepsilon_+)
\frac{}{}\right]
-
\frac{1}{\mu_1-\frac{1}{2}}
\right\}.
\label{longres1}
\end{eqnarray}
The last term in Eq.~(\ref{longres1}) is an additive constant and can be omitted
at finite $\mu_1-1/2>0$.
However, in the limit $\mu_1\to 1/2$, 
which corresponds to $\gamma\gg 1$,
the last term plays an important role.
In this limit, Eq.~(\ref{longres1}), as well as Eq.~(\ref{bSpowerlawABC})
with coefficients defined in Eq.~(\ref{ABCcoefs}) yield
\begin{equation}
\bar{S}(q,\omega)\sim \frac{m^*}{q}
\ln\left|\frac{\pi nq/m^*}{\omega-\varepsilon_+}\right|.
\label{bSlog1}
\end{equation}
In Fig.~\ref{sqwKfig}, we show
the singular dependence of $\bar{S}(q,\omega)$ on $\omega$ 
as given by Eq.~(\ref{longres1}).

Next we compare Eq.~(\ref{longres1}) to Eq.~(\ref{bSn02mv2q}) in the limit 
$\gamma\ll 1$.
We use the Bogoliubov spectrum (\ref{eqBogoSpectrum}) for the the Lieb-$1$ mode
and expand it for $q\gg m^*v$,
\begin{equation}
\varepsilon_+(q)\approx 
m^*v^2+\frac{q^2}{2m^*}.
\label{vepsexpmvs2}
\end{equation}
Substituting here $v=\sqrt{naU/m^*}$
and differentiating $\varepsilon_+(q)$ 
with respect to $n$, we obtain the characteristic scale 
$\Delta\varepsilon =n_0aU=m^*v^2$.
Thus, one should expect that 
Eqs.~(\ref{longres1}) and~(\ref{bSn02mv2q}) coincide
at $|\omega - \varepsilon_+|\ll m^*v^2$.
Taking the limit $\mu_1\to 1$ in Eq.~(\ref{longres1})
we obtain
\begin{eqnarray}
\bar{S}(q,\omega)&\simeq&
\pi\sqrt{\frac{n_0}{Ua}}
\left|\frac{1}{\omega-\varepsilon_+}\right|^{1/2}
\theta(\varepsilon_+-\omega).
\label{longres1simple}
\end{eqnarray}
On the other hand, the same result is indeed obtained
from Eq.~(\ref{bSn02mv2q}), after
expanding the r.h.s. in terms of $\omega-\varepsilon_q$,
retaining the leading order term,
and taking the limit $q\gg m^*v$.

We summarize here our result:
the exponent $\mu_1$ of the power law divergence 
of $S(q,\omega)$ at Lieb-$1$ mode,
predicted in Refs.~\onlinecite{Khodas,Imambekov1},
is reduced by $1/2$ due to averaging over the trap.

\subsubsection{Limit of strong interaction $\gamma\gg 1$ and arbitrary $q$}
\label{limofgmmagg1andarbq}

In this limit, 
the peak is asymmetric,~\cite{Khodas,Imambekov1}
\begin{eqnarray}
S(q,\omega)&\simeq& \frac{m^*}{2q}
\left|\frac{\varepsilon_+-\varepsilon_-}{\omega-\varepsilon_+}\right|^{\mu_1}
\left[
(1-\bar\nu)\theta(\omega-\varepsilon_+)\right.\nonumber\\
&&\left.
+(1+\bar\nu)\theta(\varepsilon_+-\omega)
\right].
\label{Sqwdeltastrongint}
\end{eqnarray}
Here, $0\leq \mu_1\leq 1/2$ and $0\leq \bar\nu\leq 1$ are given by  
\begin{eqnarray}
\mu_1&=&\frac{2}{\pi}\left(1-\frac{1}{\pi}\arctan Q\right)\arctan Q,\\
\bar\nu&=&\frac{1}{Q}\tan\left(\frac{\pi\mu_1}{2}\right),\label{barnuexpr}
\end{eqnarray}
with $Q=q/m^*Ua=\pi q/k_F\gamma$.
Note that here we use a slightly different parametrization
of the peak asymmetry than in Refs.~\onlinecite{Khodas,Imambekov1}.
The Lieb modes at strong interaction assume 
the expression in Eq.~(\ref{Sqwthetas}).
Equation~(\ref{Sqwdeltastrongint}) gives $S(q,\omega)$
by order of magnitude in a frequency range 
around Lieb-$1$ mode,
$|\omega-\varepsilon_+|\lesssim \varepsilon_+-\varepsilon_-$.
The equality sign in Eq.~(\ref{Sqwdeltastrongint}) holds
for the Tonks-Girardeau limit ($Q\ll 1$),
cf. Eq.~(\ref{Sqwthetas}).

\begin{figure}[t]
\includegraphics[width=0.95\columnwidth]{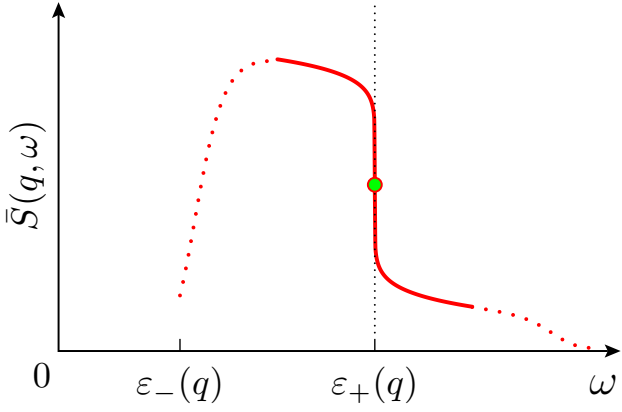}\caption{\label{sqwQfig}
Sketch of the dynamic response 
$\bar{S}(q,\omega)$ as function of $\omega$
in the regime of ``strong interaction and arbitrary $q$''
(see Sec.~\ref{limofgmmagg1andarbq}).
The response has a non-analytic point (circle)
at Lieb-$1$ mode $\varepsilon_+(q)$
(solid line) and is strongly suppressed away from $\varepsilon_+(q)$
at distances exceeding $\varepsilon_+-\varepsilon_-$ 
(dotted line).
The solid line is plotted using Eq.~(\ref{longres2})
with the choice $Q=1$, corresponding to
$\mu_1=0.375$ and $\bar{\nu}\approx 0.67$; 
the dotted line illustrates the qualitative behavior of 
$\bar{S}(q,\omega)$ away from the Lieb-$1$ mode.
The scale of the ordinate axis is arbitrary.
}
\end{figure}

We proceed in the same fashion as in Sec.~\ref{limlargeqarbint},
substituting Eq.~(\ref{Sqwdeltastrongint}) into 
Eq.~(\ref{averdef}) and using Eq.~(\ref{nofxexpand}),
and arrive at
\begin{eqnarray}
\bar{S}(q,\omega)&\sim &\frac{m^*}{2qL}
\sqrt{\frac{n_0(\varepsilon_+-\varepsilon_-)}{2|n''(0)|\Delta\varepsilon}}
\int_0^\infty\frac{d\xi}{\sqrt{\xi}}\left|\frac{1}{W+\xi}\right|^{\mu_1}\nonumber\\
&&\times
\left[
(1-\bar\nu)\theta(W+\xi) + (1+\bar\nu)\theta(-W-\xi)
\right],\nonumber\\
\label{bSintegral2}
\end{eqnarray}
where $\Delta\varepsilon=k_Fq/m^*$ and 
$W=(\omega-\varepsilon_+)/(\varepsilon_+-\varepsilon_-)$.
In Eq.~(\ref{bSintegral2}),
the parameters are taken at the trap center and the approximations
made here are the same as in Sec.~\ref{limlargeqarbint}.
In contrast to the case of Sec.~\ref{limlargeqarbint},
the integral here does not diverge at $W=0$, if $\mu_1<1/2$.
Nonetheless, Eq.~(\ref{bSintegral2}) gives the
leading order term of the non-analytic part of $\bar{S}(q,\omega)$
at Lieb-$1$ mode.
The analytic part of $\bar{S}(q,\omega)$ is not known at this order
of the asymptotic expansion in $W\ll 1$.
Performing the integration in Eq.~(\ref{bSintegral2}), we obtain
\begin{eqnarray}
\bar{S}(q,\omega)&\sim&\frac{m^*}{2qL}
\sqrt{\frac{n_0(\varepsilon_+-\varepsilon_-)}{2|n''(0)|\Delta\varepsilon}}
\left\{
\left|\frac{\varepsilon_+-\varepsilon_-}{\omega-\varepsilon_+}\right|^{\mu_1-1/2}
\right.\nonumber\\
&&\left.
\times
\frac{\sqrt{\pi}\Gamma(\mu_1-\frac{1}{2})}{\Gamma(\mu_1)}
\left[
(1-\bar\nu)\theta(\omega-\varepsilon_+)
\frac{}{}
\right.\right.\nonumber\\
&&\left.\left.
+\left(
\tan\left(\frac{\pi\mu_1}{2}\right)
-\frac{1}{Q}\right)
\theta(\varepsilon_+-\omega)
\right]\right.\nonumber\\
&&\left.
+
\frac{1-\bar\nu}{\frac{1}{2}-\mu_1}
\right\}.
\label{longres2}
\end{eqnarray}
We retained here an additive constant (last term)
to ensure proper behavior of Eq.~(\ref{longres2}) in the limit $\mu_1\to 1/2$,
in which Eq.~(\ref{longres2}) takes the form of Eq.~(\ref{bSlog1}).
In Fig.~\ref{sqwQfig}, we show
the non-analytic behavior of $\bar{S}(q,\omega)$ 
of  Eq.~(\ref{longres2})
at the Lieb-$1$ mode.

At $Q\ll 1$, Eq.~(\ref{longres2}) agrees with the 
Tonks-Girardeau limit, see Eq.~(\ref{Sqwtgroot1}).
Indeed, substituting $\mu_1\approx 2Q/\pi$ and $\bar\nu\approx 1$
in Eq.~(\ref{longres2}), we obtain
\begin{eqnarray}
\bar{S}(q,\omega)&\sim&\frac{\sqrt{2}m^*}{q}
\left|\frac{\omega-\varepsilon_+}{\Delta\varepsilon}\right|^{1/2}
\theta(\varepsilon_+-\omega).
\label{longres2simple}
\end{eqnarray}
On the other hand, expanding $\bar{S}(q,\omega)$ in Eq.~(\ref{Sqwtgroot1})
in terms of $\omega-\varepsilon_+$, 
we obtain the same result.

To conclude,
the singular behavior of $\bar{S}(q,\omega)$ at $\omega=\varepsilon_+(q)$
is owing to the weak (quadratic) dependence of $n(x)$ on $x$ in the trap center. 
More generally,
the exponent $\mu_1$ of the singular dependence changes compared to
the homogeneous case like $\mu_1\to\mu_1-1/d$, 
where $d$ is the power of the subleading order
expansion of $n(x)$ in terms of $x$ around its maximum.
Normally, 
$n''(0)\neq 0$, i.e. 
$d=2$, which is the case assumed in Eq.~(\ref{nofxexpand}).

\subsection{Dynamic structure factor of the compressible caps of the  Mott phases}
\label{SecCompRegion}

In the Mott-insulator regime the atomic cloud consists of compressible and
incompressible regions.  Here, we evaluate the density-averaged dynamic
response of a compressible region as well as its relative contribution to the
total response mentioned in Sec.~\ref{secDSFinpsenceoflattice}, see
Eq.~(\ref{Scom}).  The dynamic response of an incompressible region has been
considered in Sec.~\ref{DSFofHMPh}.

We consider the simplest case illustrated in Fig.~\ref{motttrapfig}, where the
center of the cloud is incompressible, with $na=1$, and the ends are
compressible, with $n(x)$ given in Eq.~(\ref{nofxarcsin}).  We assume $J/U\ll
1$ and agree to measure the chemical potential $\mu$ from the middle of the
first Mott band at $x=0$.  The length of one compressible region $\Delta L$ is
then given by
\begin{equation}
\frac{\Delta L}{a}=\sqrt{\frac{\mu+2J}{\epsilon_0}}-\sqrt{\frac{\mu-2J}{\epsilon_0}}\approx 
\frac{2J}{\sqrt{\mu\epsilon_0}}+{\cal O}\left((J/\mu)^3\right).
\end{equation}
We further assume that the trap potential is sufficiently weak, 
such that the compressible region is macroscopic, $\Delta L/a\gg 1$,
allowing us to use Eq.~(\ref{averdef}).

In order to calculate the dynamic response, we map the Bose-Hubbard model in
Eq.~(\ref{BH}) onto a fermionic model in the strong interaction limit $J/U\ll
1$ and at restricted site occupancy $na\leq 1$, obtaining
\begin{equation}
H=-J\sum_l\left(f_{l+1}^\dagger f_l+f_l^\dagger f_{l+1}\right)+
\sum_l (\epsilon_0l^2-\mu)f_l^\dagger f_l,
\end{equation}
where $f_l$ is a fermion annihilation operator,
$\{f_l,f_{l'}^\dagger\}=\delta_{ll'}$.  As usually, $f_l$ faithfully represents
$b_l$ in the low energy subspace ($\omega< U$) up to a non-local phase factor.
The phase factor, however, cancels out in the density operator, resulting in
$b_l^\dagger b_l = f_l^\dagger f_l$.

In the homogeneous case ($\epsilon_0=0$), the dynamic structure factor
reads\cite{Niemeijer,Mueller}
\begin{equation}
S(q,\omega)=
\frac{a^{-1}{\cal F}(q,\omega)}{\sqrt{\left(4J\sin(qa/2)\right)^2-\omega^2}},
\label{Stightbindhomogen}
\end{equation}
where
\begin{eqnarray}
&&{\cal F}(q,\omega)=\theta(\Omega_{++})\theta(-\Omega_{+-})+
\theta(\Omega_{-+})\theta(-\Omega_{--}), \nonumber\\
&&\Omega_{ss'}=\mu+2sJ\cos\left(qa/2\right)\cos\phi_0+
s'\omega/2.
\end{eqnarray}
Here, $s,s'=\pm1$ and 
\begin{equation}
\phi_0=\arcsin\left(\frac{\omega}{4J\sin(qa/2)}\right).
\end{equation}

It is convenient to relate averaging over the density profile
in Eq.~(\ref{nofxarcsin}) to an averaging over chemical potential
in expressions for the homogeneous case,
\begin{equation}
\bar{S}(q,\omega;\mu)=\frac{a}{2\Delta L}\int_{-2J}^{2J}\frac{S(q,\omega;\mu')
d\mu'}{\sqrt{\epsilon_0(\mu-\mu')}}.
\label{averagerule1}
\end{equation}
Here, we assumed that $2J<\mu<U-4J$.

Next, we average $S(q,\omega)$ in Eq.~(\ref{Stightbindhomogen}) using 
Eq.~(\ref{averagerule1}) and obtain that
$\bar{S}(q,\omega)$ is given by r.h.s. of Eq.~(\ref{Stightbindhomogen}),
but with ${\cal F}(q,\omega)\to \bar{\cal F}(q,\omega)$, where
\begin{equation}
\bar{\cal F}(q,\omega)=
\frac{\sqrt{\Omega_{++}}-\sqrt{\Omega_{+-}}+
      \sqrt{\Omega_{-+}}-\sqrt{\Omega_{--}}}
{\sqrt{\mu+2J}-\sqrt{\mu-2J}}.
\label{Favlong}
\end{equation}
Thus, we have obtained the dynamic structure factor of a compressible region,
normalized to its length.
It is convenient next to expand Eq.~(\ref{Favlong}) for $\mu\gg J$,
which corresponds to the compressible region occupying only a small
portion of the cloud, $\Delta L\ll L$.
To leading order, the dependence on $\mu$ drops out and we obtain
\begin{equation}
\bar{S}(q,\omega)=
\frac{\omega}{2Ja}
\frac{1}{\sqrt{\left(4J\sin(qa/2)\right)^2-\omega^2}},
\label{Stightbindaver1}
\end{equation}
with the support $0<\omega<4J|\sin(qa/2)|$.
We note that this result could also be obtained by making a linear
approximation for $V(x)$ over the size of the compressible
region, which is compatible with our assumption $\Delta L\ll L$.

Finally, we are in the position to obtain Eq.~(\ref{Scom})
for the contribution of one compressible region 
relative to the total length of the cloud.
We need only to multiply Eq.~(\ref{Stightbindaver1}) by $\Delta L/2L$,
where 
\begin{equation}
L=a\sqrt{\frac{\mu+2J}{\epsilon_0}}\approx a\sqrt{\frac{\mu}{\epsilon_0}}.
\end{equation}
Expressing $\mu$ via $N$ with the help of Eq.~(\ref{NEllipticE}),
$\mu=\epsilon_0N^2/4$ at $\mu\gg J$, we obtain that
$\Delta L/2L=4J/\epsilon_0N^2$ and multiplying it
in Eq.~(\ref{Stightbindaver1}) we arrive at Eq.~(\ref{Scom}).

We remark that Eq.~(\ref{Favlong}) holds also in the fully compressible regime
$-2J<\mu\leq 2J$, provided the square roots are assumed to vanish for
negative arguments, i.e.
redefining $\sqrt{x}:=\theta(x)\sqrt{x}$.

\section{Discussion}
\label{SecDSFandLA}

\subsection{Ways to measure dynamic structure factor}
\label{SecWaysToMeasure}

Bragg spectroscopy\cite{Ketterle,Steinhauer1,Steinhauer2,Richard,Muniz,Papp,Inguscio1,Inguscio2}
allows one to excite the atomic cloud with a scalar potential of the form
\begin{equation}
\varphi(x,t)={\rm Re}\left[
\varphi_0 e^{i(qx-\omega t)}
\right],
\label{varphixt}
\end{equation}
with $\omega$ and $q$ tunable
independently of each other.
The control of amplitude $\varphi_0$, ideally, is independent of 
$\omega$ and $q$.
The effective interaction of the atoms with the Bragg probe
reads
\begin{equation}
H_{\rm int}=\int dx \rho(x)\varphi(x,t).
\end{equation}
The linear response of the atomic cloud to such a perturbation
is characterized by the dynamic structure factor studied in this paper.

One way to measure the dynamic structure factor is to perform calorimetric
measurements on the system after exposing it to the driving potential
$\varphi(x,t)$ for some time. The energy gained by the system per unit time
is\cite{Nozieres}
\begin{equation}
\frac{dE}{dt}=\omega
\left|\varphi_0\right|^2 S_N(q,\omega),
\label{dEodt}
\end{equation}
where $S_N(q,\omega)\approx 2L\bar{S}(q,\omega)$ is the (unnormalized) 
dynamic structure factor of the system.

In the experiment of Ref.~\onlinecite{Inguscio1}, Bragg scattering was used to
excite the atomic cloud at a fixed momentum $q$ and a variable frequency
$\omega$ for a fixed duration of time.  The response of the atomic cloud to the
excitation was observed as a smearing of interference pattern of the atomic
density at the final stage of experiment, when the atomic cloud was left to
expand freely without confinement.  Establishing a relation between the
smearing of the interference pattern of the matter waves and the amount of
absorbed energy goes beyond the purpose of this paper.  Here, we discuss
qualitatively only the limiting case, in which the system reaches thermal
equilibrium before the trap confinement is released.  In this case, one expects
that the broadening of the interference pattern is proportional to the
temperature of the system, which has been ``heated up'' by the Bragg
excitation.  The frequency and momenta at which the system was excited enter in
the result only through Eq.~(\ref{dEodt}), and thus, one may take the quantity
$\omega S_N(q,\omega)$ as a measure for the calorimetric response.

\subsection{$f$-sum rules in the presence of optical lattice}
\label{SecfSumRules}

The response $\omega S_N(q,\omega)$ features several
properties which we would like to mention:

\noindent{\em 1. Complete $f$-sum rule.} Regardless of the interaction strength,
trap potential, and optical lattice, the particle number conservation
for atoms imposes the constraint
\begin{equation}
\int_0^{\infty}\frac{d\omega}{2\pi}\omega S_N(q,\omega)=\frac{Nq^2}{2m},
\label{fsumrule}
\end{equation}
where $m$ is the bare atomic mass.
According to Eq.~(\ref{fsumrule}), the net spectral weight
of the calorimetric response is conserved and can change only 
with particle number $N$ and probe momentum $q$.

\noindent{\em 2. Partial $f$-sum rule I.}
A similar sum rule to the one in Eq.~(\ref{fsumrule}) can be
formulated in the continuum limit of the optical lattice problem.
For the cases considered in Sec.~\ref{Bogolimit}, \ref{TGlimit}, and~\ref{ArbInt},
the partial $f$-sum rule reads
\begin{equation}
\int_0^{4J}\frac{d\omega}{2\pi}\omega S_N(q,\omega)=\frac{Nq^2}{2m^*}.
\label{partIfsumrule}
\end{equation}
Equation~(\ref{partIfsumrule}) is valid for $q\ll \pi/a$ and
provided that $\bar\mu, U \ll 4J$.
This allows us to choose
the upper limit of integration to be on the order of the band width 
$4J$ and to use the effective mass $m^*$.
The remaining spectral weight $\propto(1/m-1/m^*)$ is distributed between
higher energy bands, at frequencies above the recoil energy 
$E_R=\pi^2/2ma^2$.
The interaction affects the shape of the integrand in Eq.~(\ref{partIfsumrule}),
but does not affect the net weight as long as interband matrix elements of the
interaction can be neglected.
One may check that $S_N(q,\omega)$ evaluated with the help of 
Eqs.~(\ref{bSmqsqrt}), (\ref{bSn02mv2q}), and~(\ref{Sqwtgroot1})
satisfies the rule in Eq.~(\ref{partIfsumrule}).

\noindent{\em 3. Partial $f$-sum rule II.}
Within the Bose-Hubbard model,
following the standard derivation of $f$-sum rule,\cite{Nozieres}
one finds
\begin{equation}
\int_0^{\infty}\frac{d\omega}{2\pi}\omega S_N(q,\omega)=-2\sin^2(qa/2)
\left\langle H_1\right\rangle
\label{BHfsumrule}
\end{equation}
where $H_1$ is the tunneling Hamiltonian in Eq.~(\ref{H1})
and average $\langle\dots\rangle$ is taken over the ground state.
The upper limit of integration in Eq.~(\ref{BHfsumrule})
is formally infinity for the Bose-Hubbard model.
For an optical lattice in the tight-binding limit $J,U\ll E_R$,
the upper limit of integration can be replaced by $E_R$.
Equation (\ref{BHfsumrule}) can be considered as a generalization
of Eq.~(\ref{partIfsumrule}), to which it reduces in the limit
$q\ll\pi/a$ and $\bar\mu, U\ll 4J$.
We remark that the following inequality 
$-\left\langle H_1\right\rangle\leq 2JN$
holds for arbitrary occupation $na<1$ in the Bose-Hubbard model.
Therefore, the spectral weight is bounded from above as follows
\begin{equation}
\int_0^{E_R}\frac{d\omega}{2\pi}\omega S_N(q,\omega)\leq 4JN\sin^2(qa/2).
\label{BHfsumrulebound}
\end{equation}
The latter equation shows that
the dynamic response in the low frequency range $\omega\in[0,E_R]$
is suppressed at least as $\propto J$.
A suppression that is only proportional to $J$ (or $1/m^*$) is due to the
optical lattice and is present also for non-interacting atoms.
The interaction leads to an additional suppression when entering into the Mott regime.

\noindent{\em 4. Suppression of dynamic response in the Mott regime.}
In the Mott phase the tunneling is quenched to first order of $J$,
i.e. $\left\langle H_1\right\rangle={\cal O}(J^2)$.
At the second order of $J$, we find 
$\left\langle H_1\right\rangle=2N\Delta{\cal E}_0$, where $\Delta{\cal E}_0<0$
is the correction to the ground state energy per atom 
due to tunneling.
Thus, for the Mott phase we have according to Eq.~(\ref{BHfsumrule})
\begin{equation}
\int_0^{E_R}\frac{d\omega}{2\pi}\omega S_{N}(q,\omega)=
4|\Delta{\cal E}_0|
N\sin^2(qa/2).
\label{BHMottspcwght}
\end{equation}
Evaluating $\Delta{\cal E}_0$ by perturbation theory, 
we find $\Delta{\cal E}_0=-2(p+1)J^2/U$, where we kept
for generality an arbitrary Mott occupation number $p$.
One may check that $S_N(q,\omega)$ evaluated with the help of
Eq.~(\ref{Sincsemicirclepp})
satisfies the rule in Eq.~(\ref{BHMottspcwght}).
Since $|\Delta{\cal E}_0|\ll J$, a natural question arises:
How does the suppression of dynamic response occur with entering into the Mott regime?
To answer this question, it is instructive to consider
a homogeneous compressible liquid in the limit $J/U\ll 1$.
Considering $na< 1$, we obtain from Eq.~(\ref{BHfsumrule})
\begin{equation}
\int_0^{E_R}\frac{d\omega}{2\pi}\omega S_{N}(q,\omega)=
4JN\sin^2(qa/2)
\frac{\sin(\pi n a)}{\pi n a}.
\label{BHMottcompliq}
\end{equation}
Here,
the factor $\sin(\pi n a)/\pi n a$, which vanishes at $na\to 1$,
has at its origin the phase space restriction imposed by the
interaction.
This factor governs the transition between
the upper bound in 
Eq.~(\ref{BHfsumrulebound}) and Eq.~(\ref{BHMottspcwght})
with entering into the Mott regime.
Finally, we remark that Eq.~(\ref{BHMottcompliq}) is generalized
to larger occupations 
$p\leq na<p+1$ by replacing $J\to (p+1)J$ and $na\to na-p$, where
$p$ is a positive integer.

\subsection{Relation to the experiment of Ref.~\onlinecite{Inguscio1}}
\label{RelToExperiment}

\begin{figure}[t]
\includegraphics[width=0.95\columnwidth]{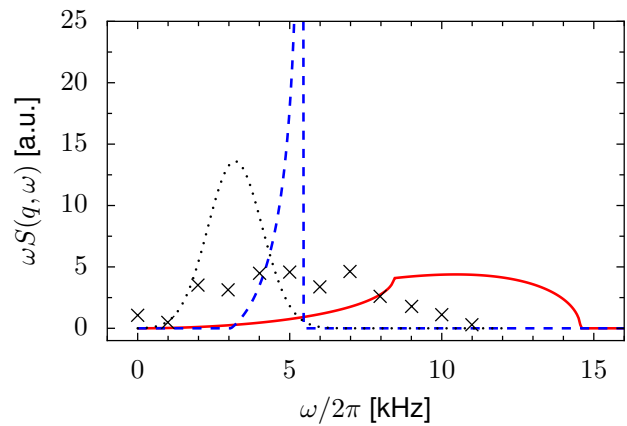}\caption{\label{wsqwclfig}
    Attempted comparison of a theory for a single tube at $T=0$ with
    experiment for the case without an optical lattice.  Experiment:
    The data points ($\times$) are taken from Fig.~1(c) of
    Ref.~\onlinecite{Inguscio1}.  Theory: Several limiting cases are
    shown.  The solid line shows the dynamic response in the
    Tonks-Girardeau limit ($\gamma\to\infty$) given by
    Eq.~(\ref{bSmqsqrt}).  The dashed line shows the response
    evaluated in the Bogoliubov limit, see Eq.~(\ref{bSn02mv2q}), at
    the value of $mv^2\approx 3.2\,{\rm kHz}$.  The dotted line shows
    the response of non-interacting bosons ($\gamma=0$) given by
    Eq.~(\ref{SqwnonintdirectGauss}).  }
\end{figure}

In this section, we turn to a discussion of the dynamic response measurement
reported in Ref.~\onlinecite{Inguscio1}.  At first, we consider the case
without optical lattice, namely the case of Fig.~1(c) of
Ref.~\onlinecite{Inguscio1}.  The dynamic response is a single broad peak
centered at $\omega/2\pi\approx 5\,{\rm kHz}$ with a FWHM of about $7\,{\rm
kHz}$, see Fig.~\ref{wsqwclfig}.  The response comes from an array of 1D tubes
loaded with ${}^{87}{\rm Rb}$ atoms, with $N\approx 250$ atoms in the central
tube and with a longitudinal trap frequency of $\omega_0/2\pi\approx 43\,{\rm
Hz}$.  The system is probed at $q\approx 7.3\times 10^{6}\,{\rm m}^{-1}$.  In
Fig.~\ref{wsqwclfig}, we compare the predictions of our theory against the
experimental data.  The solid curve shows the dynamic response in the
Tonks-Girardeau limit (see Sec.~\ref{TGlimit}) and is plotted using
Eq.~(\ref{bSmqsqrt}) with $\pi n_0\equiv k_F=\sqrt{2m\bar{\mu}}$ and
$\bar{\mu}=N\omega_0$.  The dashed curve in Fig.~\ref{wsqwclfig} shows the
dynamic response in the Bogoliubov limit (see Sec.~\ref{Bogolimit}) and is
plotted using Eq.~(\ref{bSn02mv2q}) with $mv^2\approx 3.2\,{\rm kHz}$. We
explain below how we determine the value of the interaction energy $mv^2$ in
this limit. The dotted line in Fig.~\ref{wsqwclfig} shows the dynamic response
in the limit of non-interacting bosons (see
Appendix~\ref{AppFiniteSizeEffects}) and is plotted using
Eq.~(\ref{SqwnonintdirectGauss}). Unlike the previous two results, the result
shown by the dotted line takes into account the ``quantumness'' of the trap
potential, see Appendix~\ref{AppFiniteSizeEffects}. Note that we used no fit
parameters in Fig.~\ref{wsqwclfig}, except for rescaling all curves to have the
same area to obey the $f$-sum rule.

\begin{figure}[t]
\includegraphics[width=0.95\columnwidth]{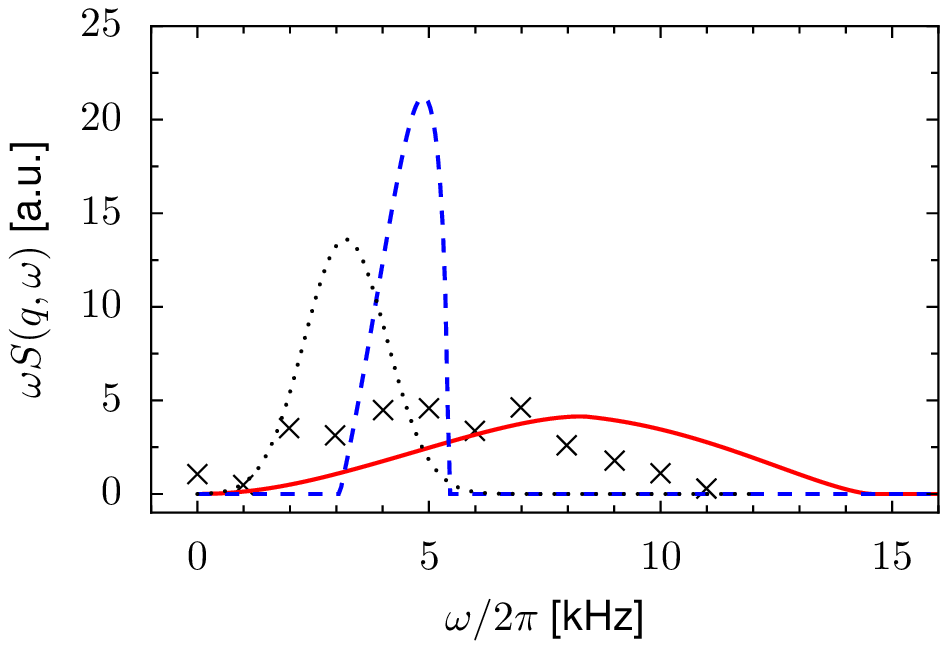}\caption{\label{wsqwcl4fig}
The same as in Fig.~\ref{wsqwclfig}, but with averaging over a 2D array of 1D
bosonic systems with different particle number $N$, see
Appendix~\ref{AppAvOverTubes}.  The solid line  (Tonks-Girardeau limit) is
plotted using Eq.~(\ref{eqSNTGNAVER}) and the dashed line (Bogoliubov limit)
using Eq.~(\ref{SNunaverbogoAVR}).  The dotted line (non-interacting bosons) is
insensitive to averaging over $N$ at zero temperature.
}
\end{figure}

The value of interaction energy $mv^2$ in the Bogoliubov limit was determined
as follows.  Using expressions for weak interaction, we relate $\bar \mu$ to
$N$ and $g$, obtaining
\begin{equation}
\label{eq:mubog}
\bar\mu\equiv mv^2=\left[\frac{3}{4\sqrt{2}}\sqrt{m}g\omega_0N\right]^{2/3}.
\end{equation}
The coupling constant $g$ can be extracted from the experimental data in the
Mott regime.  In particular, we consider Figs.~1(f)-(h) of
Ref.~\onlinecite{Inguscio1} and single out a series of peaks which do not shift
towards lower energies with increasing the height of the optical lattice
potential.  These peaks are attributed to the dynamic response occurring at
multiples of $U$ in the Mott regime.  In the case of
Ref.~\onlinecite{Inguscio1}, we were able to identify the peaks occurring at
$\omega/2\pi\approx 2\,{\rm kHz}$ and $4\,{\rm kHz}$ with the expected dynamic
response at $U$ and $2U$, respectively.  Our identification of peaks agrees
with the one reported in Ref.~\onlinecite{Inguscio1}.  Having the value of
$U/2\pi= 2\,{\rm kHz}$ at a height of the lattice potential of approximately
$13E_R$ (see Fig.~1(g) of Ref.~\onlinecite{Inguscio1}), we evaluate $g$ using
tight-binding theory,
\begin{equation}
\label{eq:g}
g=\sqrt{\frac{2}{\pi}}s^{-1/4} Ua,
\end{equation}
where $s$ is the potential height ($s=13$) measured in units of
$E_R=\pi^2/2ma^2$.  The lattice potential is assumed to be
$V(x)=sE_R\sin^2(\pi x/a)$ and the lattice constant in the experiment is
$a=415\,{\rm nm}$.  With this information, we obtain the value of $mv^2\approx
3.2\,{\rm kHz}$.  We note that this value of interaction energy corresponds to
a value of the dimensionless interaction strength of $\gamma\approx 0.32$.
This agrees by order of magnitude with the value of $\gamma\approx 0.6$
reported in Ref.~\onlinecite{Inguscio1}. The discrepancy may stem from a
different definition of $\gamma$, which involves averaging
over an ensemble of 1D systems.\cite{note25}
Our definition of $\gamma$ refers to a single 1D system and it is as follows
\begin{equation} 
\gamma=\frac{mg}{n_0},
\end{equation} 
where $n_0$ is the density in the trap center. With this
definition, one can estimate $\gamma$ as 
\begin{eqnarray}
\gamma&=&\min\left\{\gamma_<,\gamma_>\right\},\\
\gamma_<&=&\left[\frac{4\sqrt{2}}{3}\frac{m g^2}{\omega_0N}\right]^{2/3},\\
\gamma_>&=&\pi g\sqrt{\frac{m}{2\omega_0N}},
\end{eqnarray}
where $\gamma_<$ ($\gamma_>$) is $\gamma$ obtained using the expression for
$n_0$ in the Bogoliubov (Tonks-Girardeau) limit.

\begin{figure}[t]
\includegraphics[width=0.95\columnwidth]{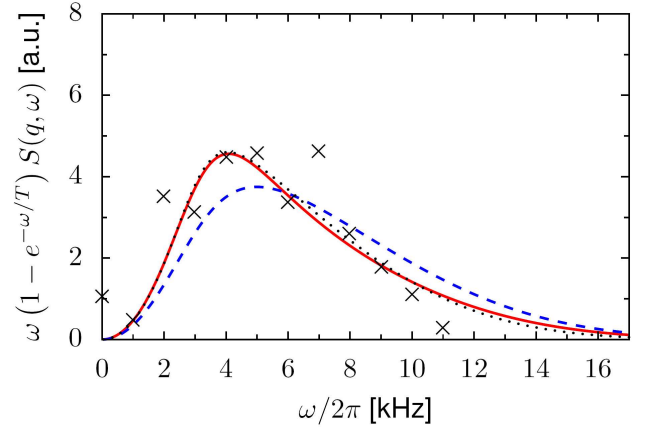}\caption{\label{wsqwtempfig}
    Dynamic response of free bosons at a finite temperature, see
    Appendix~\ref{AppDynResFinTemp}.  Data points ($\times$) are the
    same as in Figs.~\ref{wsqwclfig} and~\ref{wsqwcl4fig} and are
    taken from Ref.~\onlinecite{Inguscio1}. Solid line shows the
    dynamic response of non-interacting bosons in a single 1D trap at
    $T=4.2\,{\rm kHz}$ (least-squeares fit) and $N=250$. Dashed line
    shows the response of an ensemble of such 1D systems with $N$
    distributed according to Eq.~(\ref{PofM}) and with $N_{\rm
      max}=250$. Temperature is kept at $T\approx 4.2\,{\rm kHz}$ to
    illustrate the broadening of the peak due to ensemble averaging.
    Dotted line (close to solid line) shows the ensemble-averaged
    response at $T\approx 3.3\,{\rm kHz}$ (least-squeares fit). All
    curves are rescaled to have equal areas in accordance with the
    $f$-sum rule.  }
\end{figure}

Next, we return to the comparison made in Fig.~\ref{wsqwclfig}.  We note that
the measurement data in Fig.~\ref{wsqwclfig} display a much broader maximum
than the one predicted by the weak interaction (dashed) or free-boson (dotted)
theory.  The maximum predicted by the strong-interaction theory (solid) is
sufficiently broad, but it is offset by a large energy ($\approx 5\,{\rm kHz}$)
away from the position of the measured data.  In order to gain an insight into
the origin of the discrepancy we verify two hypotheses: (i) averaging over the
array of 1D traps is responsible for the broadening of the peak and (ii) finite
temperature is responsible for the broadening of the peak.  We carry out the
averaging over the 2D array of 1D systems in Appendix~\ref{AppAvOverTubes}. We
take into account the fact that different 1D systems may have different
particle number $N$, using the probability distribution $P(N)$ in
Eq.~(\ref{PofM}) taken form Ref.~\onlinecite{Bloch}. The distribution $P(N)$
depends only on the particle number $N_{\rm max}$ in the central tube of the 2D
array.  We take $N_{\rm max}=250$ and plot our results of
Eqs.~(\ref{eqSNTGNAVER}) and~(\ref{SNunaverbogoAVR}) in Fig.~\ref{wsqwcl4fig}.
Averaging over the 2D array modifies the analytic properties of
$S_{N}(q,\omega)$ at the Lieb modes, see Appendix~\ref{AppAvOverTubes}. The
peak shape has changed significantly for the Bogoliubov  and Tonks-Girardeau
limits as compared to the case of a single 1D system. Nonetheless, the width
of the peaks is unaffected by the averaging. This can be easily understood by
recalling that the 1D system with the largest particle number $N$ has the
largest dynamic response (see $f$-sum rules in Sec.~\ref{SecfSumRules}).  Thus,
the dynamic response in a 2D array of tubes is dominated by the tubes close to
the central tube, which have about the same number of atoms $N\sim N_{\rm
max}$. We conclude that averaging over $N$ in the 2D array cannot explain the
discrepancy seen in Fig.~\ref{wsqwclfig}.

Secondly, we verify hypothesis (ii) and analyze the effect of finite
temperature in Appendix~\ref{AppDynResFinTemp}. 

We find that already noninteracting bosons at high temperature would
yield a spectral function with a width of the order to the one
measured in the experiments, see Fig.~\ref{wsqwtempfig}. This fact is
not surprising, since the momentum $q$ at which the system is probed
is sufficiently large (a) to cause a sizable broadening of the
free-boson peak (see dotted line in Fig.~\ref{wsqwclfig}) and (b) to
increase the exponent $\mu_1$ of the power-law singularity of
$S(q,\omega)$ at Lieb-$1$ mode in the homogeneous Lieb-Liniger model.
Despite the fact that effect (b) refers to the homogeneous case, the
physical picture behind it applies to any system: interaction effects
are less pronounced at large momenta of the probe. In particular, for
the Lieb-Liniger model at $\gamma\neq 0$ the exponent $\mu_1$
increases monotonically from $\mu_1=0$ at $q=0$ to $\mu_1=1-1/2K$ at
$q\gg q_0$, where $q_0\sim \max\left(mv,mg\right)$ is the momentum
scale defined in Eq.~(\ref{q0expl}). At $q\gg q_0$, a quantum
$(q,\omega)$ of the probe excites single bosons projectively rather
than exciting many bosons collectively via processes involving
interaction. For reference, we note~\cite{unpub} that at $\gamma=0.6$
the exponent $\mu_1\approx 0.2$ at $q=0.1\pi n_0$ and $\mu_1\approx
0.8$ at $q=\pi n_0$. Similarly, effect (a) becomes pronounced with
increasing $q$; the width of the peak scales as $\delta\omega\sim
q/m\lambda$, see Appendix~\ref{AppFiniteSizeEffects}, making any
interaction-induced broadening difficult to observe at large momenta
($q/m\lambda\gg mv^2$). A finite temperature complicates the
distinction between the interaction and quantum finite--size effects. In
Fig.~\ref{wsqwtempfig}, the solid line shows the fit to the
experimental data of the free-boson theory for a single 1D trap, see
Appendix~\ref{AppDynResFinTemp}; the best fit is obtained at
$T=4.2\,{\rm kHz}$. A characteristic sharp peak of width
$\delta\omega\sim q/m\lambda$ with long tails extending over
$\delta\omega\sim q\sqrt{T/m}$ is indicative of the regime of high
temperatures, $\omega_0\ll T\ll \omega_0N/\ln N$. In the very-high
temperature regime, $T\gg \omega_0N/\ln N$, the sharp peak is absent
and the dynamic response is a single broad peak.  In
Fig.~\ref{wsqwtempfig}, the temperature is intermediate between
``high'' and ``very high'', $T\sim \omega_0N/\ln N$, which explains
the peculiar shape of the peak. In contrast to the case of $T=0$ (cf.
dotted line in Fig.~\ref{wsqwcl4fig}), ensemble averaging affects the
free-boson dynamic response at $T\neq 0$. Using the probability
distribution in Eq.~(\ref{PofM}), we carry out the averaging over $N$
and show the result in Fig.~\ref{wsqwtempfig} (dashed line); note that
$T$ for the dashed line was kept at the same value as for the solid
line. Not unlike the results of Fig.~\ref{wsqwcl4fig}, ensemble
averaging does not lead to substantial broadening of the peak (width
increses by factor order unity). Taking into account the ensemble
averaging in our fit, we obtain a smaller value of temperature,
$T\approx 3.3\,{\rm kHz}$, see dotted line in Fig.~\ref{wsqwtempfig}.
Since we neglected interactions, this value of temperature is an upper
bound on the temperature in the experiment.

In the following, we briefly discuss the case with an optical lattice.  At
sufficiently strong binding of the optical lattice, the atomic cloud enters
into the Mott regime, see Figs.~1(f)-(h) of Ref.~\onlinecite{Inguscio1}.  As
mentioned  above, we can identify here several peaks which do not shift to
lower frequencies with increasing the optical lattice amplitude.  These peaks,
in particular the two peaks occurring at frequencies $\approx 2\,{\rm kHz}$ and
$\approx 4\,{\rm kHz}$, can be attributed to excitations of the Mott phase
occurring at frequencies $U$ and $2U$, respectively.  At even lower
frequencies, the peaks are most probably associated with the response of
compressible domains, see Sec.~\ref{SecCompRegion}, at $\omega\lesssim 4(p+1)J$,
with $p$ being the underlying occupation number, see Eq.~(\ref{occupnump}).  At
low temperatures ($T\ll U$), the main contribution to the peak at
$\omega\approx U$ comes from the incompressible phases.  According to
Eqs.~(\ref{Sincsemicircle}) and~(\ref{Scom}), the contribution of the
compressible phases at $\omega\sim J$ and incompressible ones at $\omega\approx
U$ into the $f$-sum rule are of the same order\cite{note2} at $q\sim\pi/a$ and
$N\sim N_0$.  This is in a qualitative agreement with the experiment, in spite
of the fact that probably $T\sim U$ in the experiment.  The spectral weight of
the peak at $\omega\approx 2U$ is by $\sim(J/U)^2$ smaller than that of the
$\omega\approx U$ peak, see Eq.~(\ref{manymotts}).  The tendency of the
decrease of $\omega\approx 2U$ peak weight is seen in Figs.~1(f) and~1(g) of
Ref.~\onlinecite{Inguscio1}.

The dynamic response measured in Ref.~\onlinecite{Inguscio1} features a strong
suppression in the range of frequencies $0<\omega<E_R$ as the system enters
into the Mott regime.  Such a suppression is to be expected from the $f$-sum
rule in Eq.~(\ref{BHfsumrule}).

\section*{Conclusion}
We considered the effect of a smooth trap potential on
the momentum-resolved inelastic light scattering off a one-dimensional
interacting atomic gas. Cases with and without optical lattice
superimposed on the trap potential are analysed in detail. The
singularities in the scattering cross-section associated with the
excitation of collective modes persist, though in a modified form,
even upon the averaging over the varying along the trap gas density.
Rounding of the singularities due to the confinement occurs only at
the smallest momentum transfers $q$ of the order $q\sim
1/\sqrt{m\omega_0}$, where $m$ and $\omega_0$ are, respectively, the
mass and oscillation frequency for an atom in a trap. Another source
of smearing, apparently dominant in the current
experiments~\cite{Inguscio1}, is a finite temperature of the gas.

\acknowledgements We thank D. Clement for exchanges about the
experiment of Ref.~\cite{Inguscio1}.  We acknowledge support from the
USA DOE Grant No.~DE-FG02-08ER46482 and the Nanosciences Foundation at
Grenoble, France.  AM acknowledges funding from the MIDAS-STREP
project.  VNG acknowledges support from DFG (SFB 631, SFB-TR12).

\appendix
\section{quantum finite-size effect}
\label{AppFiniteSizeEffects}

In the bulk of the paper, we consider the effect of 
a varying atomic density $n(x)$ on the dynamic response
by employing Eq.~(\ref{averdef}).
There exists a competing effect to the one accounted for by
Eq.~(\ref{averdef}). 
The competing effect stems from the finite size of the atomic could
and it is most vividly seen in the
non-interacting limit. 
In that limit,  the averaging
in Eq.~(\ref{averdef}) has no effect, because
the Lieb-$1$ mode is trivially equal to
$\varepsilon_+(q)=q^2/2m$ and thus is independent of $n$.
In what follows, we consider the example of non-interacting bosons
and determine the magnitude of the competing effect.

For non-interacting bosons,
the dynamic structure factor can be calculated directly,
\begin{equation}
S_N(q,\omega)=2\pi N\sum_{\ell=1}^{\infty}
\frac{\left(q\lambda\right)^{2\ell}}{2^{\ell}\ell!}
e^{-\frac{q^2\lambda^2}{2}}
\delta(\omega-\ell\omega_0),
\label{Sqwnonintdirect}
\end{equation}
where $\lambda=1/\sqrt{m\omega_0}$ is the quantum length scale of the trap
and $\omega_0$ is the trap frequency related to $\epsilon_0$
in Eq.~(\ref{conf}) by $\omega_0=(1/a)\sqrt{2\epsilon_0/m}$.
For simplicity, we consider the case without an optical lattice.
It is convenient to replace the sum over the discrete $\delta$-peaks
in Eq.~(\ref{Sqwnonintdirect}) by a continuous envelope function.
Taking the continuum limit of $\ell$ in Eq.~(\ref{Sqwnonintdirect}), we obtain
\begin{equation}
S_N(q,\omega)=N
\sqrt{\frac{2\pi}{\omega\omega_0}}
\left(
\frac{q^2}{2m\omega}
\right)^{\omega/\omega_0}
\exp\left(\frac{\omega-q^2/2m}{\omega_0}\right).
\label{Sqwnonintdirectcont}
\end{equation}
Furthermore, since we are interested in small $\omega_0$, we
may take the limit $q^2/m\omega_0\gg 1$ in Eq.~(\ref{Sqwnonintdirectcont})
and obtain for the leading order asymptotic term
\begin{equation}
S_N(q,\omega)=N
\sqrt{\frac{2\pi}{\omega\omega_0}}
\exp\left[-
\frac{\left(\omega-q^2/2m\right)^2}{\omega_0q^2/m}
\right].
\label{SqwnonintdirectGauss}
\end{equation}
In the limit $\omega_0\to 0$, Eq.~(\ref{SqwnonintdirectGauss}) becomes
$S_N(q,\omega)=2\pi N\delta(\omega-q^2/2m)$ as expected for the case 
without a trap.

We summarize the previous paragraph as follows.
The $\delta$-peak in $S(q,\omega)$,
occurring at frequency $\varepsilon_+(q)=q^2/2m$
for free non-interacting bosons,
is broadened by $\delta\omega=q/m\lambda$, 
due to the ``quantumness'' of the trap potential.
The width $\delta\omega$ can be rewritten in a more general way
\begin{equation}
\delta\omega=\frac{\partial \varepsilon_+}{\partial q}\delta q,
\label{width1dq}
\end{equation}
which holds also in the presence of an optical lattice.
Here, $\delta q$ is the momentum uncertainty
in the ground state due to the zero-point motion in the trap.
For the case of optical lattice, $\varepsilon_+(q)$ is the
band dispersion relation and $\delta q\sim 1/\lambda^*$, where
$\lambda^*$ is the characteristic extension of the ground state wave function.
For $\omega_0\ll 4J$, the length $\lambda^*$ is given by the
expressions above, with $m\to m^*$.

For interacting bosons, both the momentum uncertainty and the
density variation affect $S(q,\omega)$.
Equation~(\ref{Sqwnonintdirect}) gives the limiting behavior
of the dynamic structure factor
for non-interacting bosons (or extremely weakly interacting, $\bar{\mu}\ll \omega_0$)
and
Eq.~(\ref{averdef}) gives the limiting behavior 
for a nearly classical (``soft'') trap potential.
The crossover between the two limiting cases has
a governing parameter, which 
involves the interaction strength, curvature of the
trap potential, and the momentum $q$ at which the system is probed.
To determine this parameter, we add to
Eq.~(\ref{width1dq}) a term representing the smearing due to the
density averaging employed in Eq.~(\ref{averdef}),
\begin{equation}
\delta\omega=\frac{\partial \varepsilon_+}{\partial n}\delta n+
\frac{\partial \varepsilon_+}{\partial q}\delta q,
\label{width2dqdn}
\end{equation}
where $\delta n$ is on the order of $n_0$ and represents the density variation in the trap.
The governing parameter is obtained
by taking the ratio of the two terms in Eq.~(\ref{width2dqdn}).
As an example, let us consider the limit of weak interaction and a sufficiently large $q$,
such that we may use Eq.~(\ref{vepsexpmvs2}) for the Lieb-$1$ mode.
We obtain from Eq.~(\ref{width2dqdn}) that 
$\delta\omega=\bar\mu+q/m^*\lambda^*$.
Thus, the density averaging prescription of Eq.~(\ref{averdef}) is valid for
\begin{equation} 
q\ll \bar\mu m^* \lambda^*.
\label{qllmumlam}
\end{equation}
This condition is compatible with the requirement
$q\gg 1/\lambda$, see Sec.~\ref{AvDyStFa},
if $\bar\mu\gg \omega_0$.
Similar (but not identical) 
criteria as in Eq.~(\ref{qllmumlam})
can be worked out 
using Eq.~(\ref{width2dqdn})
for each limiting case in which $\varepsilon_+(q)$ is known.

A different (though related) 
finite-size effect is the discreteness of energy levels in the trap
given by the trap frequency $\omega_0$.
This discreteness introduces a ``coarse graining'' of the dynamic response along the frequency axis.
In practice, $\omega_0$ is negligibly small.
In the main text of this paper, we assumed that both $\omega_0$ and the temperature are the smallest 
energy scales
in the problem and neither of them is resolved on the scale of $\omega$.

\section{Averaging over an array of 1D systems}
\label{AppAvOverTubes}

Here, we consider an ensemble of 1D systems with different number of atoms $N$.
We average our results with the following probability distribution\cite{Bloch}
\begin{equation}
P(N)=\frac{2}{3}\frac{1}{N_{\rm max}^{2/3}N^{1/3}}, \quad\quad N\leq N_{\rm max},
\label{PofM}
\end{equation}
where $N_{\rm max}$ is the maximal number of atoms per 1D system found in the
ensemble.
The probability distribution in Eq.~(\ref{PofM}) describes a 
situation typically realized in the experiment.\cite{Bloch,Weiss,Inguscio1,Inguscio2} 
Namely, a large number ($10^2$--$10^3$) of 1D systems form out of
a 3D Bose-Einstein condensate after an exponential ramp up of
a 2D optical lattice potential.
The distribution of atoms over the array of 1D systems is
obtained as a snapshot of the original 3D density profile
(assuming that the atoms had no time to migrate between the 1D systems).
For a 3D Bose-Einstein condensate in a spherically-symmetric harmonic trap, 
the number of atoms per 1D system at position $(i,j)$ in the 2D array
is given by\cite{Bloch,note3}
\begin{equation}
N_{i,j}=N_{\rm max}\left[1-\frac{2\pi N_{\rm max}}{5N_{\rm 3D}}\left(i^2+j^2\right)\right]^{3/2},
\label{Nij}
\end{equation}
where $N_{\rm 3D}$ is the total number of atoms loaded into the 3D trap.
The distribution in Eq.~(\ref{PofM}) is readily obtained
from Eq.~(\ref{Nij}), after
counting the number of 1D systems with $N_{ij}=N$ for a large 2D array.

We start with the Bogoliubov limit analyzed for 
a single 1D system in Sec.~\ref{Bogolimit}.
We express the unnormalized dynamic structure factor 
$S_N(q,\omega)=2L\bar{S}(q,\omega)$ explicitly though the number of 
atoms $N$ in the 1D system.
Using Eqs.~(\ref{bgas}), (\ref{bSn02mv2q}), and (\ref{eqofstatebosons})
as well as $L=a\sqrt{\bar\mu/\epsilon_0}$ and $\epsilon_0=m\omega_0^2a^2/2$,
we obtain
\begin{equation}
S_N(q,\omega)=\frac{2\sqrt{2}}{\omega_0g}\frac{\pi}{q}
\frac{\omega^2-q^2/2m}{\sqrt{\varepsilon_+(q)-\omega^2}}
\label{SNunaverbogo}
\end{equation}
for $q^2/2m <\omega<\varepsilon_+(q)$, and $S_N(q,\omega)=0$ otherwise.
The structure factor in Eq.~(\ref{SNunaverbogo})
depends on $N$ only through the speed of sound $v$ entering
the expression for
$\varepsilon_+(q)$ in Eq.~(\ref{eqBogoSpectrum}).
The speed of sound $v$ depends on $N$ as follows
\begin{equation}
v=\left(\frac{3\omega_0g}{4\sqrt{2}m}\right)^{1/3}N^{1/3}
\end{equation}
We average $S_N(q,\omega)$ in Eq.~(\ref{SNunaverbogo})
over $N$ with the probability distribution in Eq.~(\ref{PofM})
and obtain
\begin{equation}
\bar{S}_N(q,\omega)
=\frac{2^{5/2}\pi}{q^3v^2\omega_0g}\left(\omega^2-\frac{q^2}{2m}\right)
\sqrt{\varepsilon_+(q)-\omega^2}
\label{SNunaverbogoAVR}
\end{equation}
for $q^2/2m <\omega<\varepsilon_+(q)$, and $\bar{S}_N(q,\omega)=0$ otherwise.
Here, we replaced $N_{\rm max}\to N$ after performing the averaging.
We conclude that 
averaging over an array of 1D systems modifies the analytic behavior
of $S_N(q,\omega)$.
The one-over-square-root singularity present in $S_N(q,\omega)$
before the averaging is replaced by a square-root non-analyticity
after the averaging.
The averaging does not change the width of the peak of $S_N(q,\omega)$
versus $\omega$, provided we 
agree to compare the averaged response against 
the response of a 1D system with $N=N_{\rm max}$.

Next, we turn to the Tonks-Girardeau limit analyzed for
a single 1D system in Sec.~\ref{TGlimit}.
Using Eqs.~(\ref{fgasn0a}), (\ref{bSmqsqrt}), and~(\ref{Ncontferm}),
we obtain
\begin{eqnarray}
S_N(q,\omega) &=&
\frac{2^{3/2}m^{1/2}}{q\omega_0^{1/2}}
\left[
\sqrt{N-\frac{\left(\omega-q^2/2m\right)^2}{2\omega_0q^2/m}}
\right.\nonumber\\
&&\left.-
\sqrt{N-\frac{\left(\omega+q^2/2m\right)^2}{2\omega_0q^2/m}}
\right],
\label{eqSNTGN}
\end{eqnarray}
where the square roots are redefined as $\sqrt{x}:=\theta(x)\sqrt{x}$.
Averaging the dynamic response in Eq.~(\ref{eqSNTGN}) over $N$
with the probability distribution in Eq.~(\ref{PofM}),
we obtain
\begin{eqnarray}
\bar{S}_N(q,\omega) &=&
\frac{2^{3/2}m^{1/2}N^{1/2}}{q\omega_0^{1/2}}
\left[
f\left(\frac{\left(\omega-q^2/2m\right)^2}{2\omega_0Nq^2/m}\right)
\right.\nonumber\\
&&\left.-
f\left(\frac{\left(\omega+q^2/2m\right)^2}{2\omega_0Nq^2/m}\right)
\right],
\label{eqSNTGNAVER}
\end{eqnarray}
where the function $f(x)$ is defined as follows
\begin{equation}
f(x)=\frac{4}{7}x\sqrt{1-x}\left[\frac{1}{x}-{}_2{\rm F}_1\left(\frac{1}{3},1,\frac{3}{2},1-x\right)\right],
\label{fofxfunct}
\end{equation}
with ${}_2{\rm F}_1(\alpha,\beta,\gamma,z)$ being the hypergeometric function
and the square root (re)defined as before.
Not unlike the previous case, 
the averaging does not change the peak width in $S_N(q,\omega)$ versus $\omega$.
The non-analytic properties of $S_N(q,\omega)$ are altered by the
averaging, making the non-analyticities at $\omega=\varepsilon_\pm(q)$
less pronounced.
We note that $f(x)\approx (4/9)(1-x)^{3/2}$ at $x\to 1$.

\section{Dynamic response at finite temperatures}
\label{AppDynResFinTemp}

In the bulk of the paper, we were considering dynamic response at zero
temperature.  Many of the peculiar features of $S(q,\omega)$ vanish
gradually with increasing the temperature.  The temperature can
usually be regarded as the low energy cutoff limiting the
observability of the singular and non-analytic behaviors of
$S(q,\omega)$.  A generalization of our theory to finite temperature
is possible, although it goes beyond the purpose of this paper.  In
this paper, our goal was to establish the regimes in which the
power-law singularities of $S(q,\omega)$ derived for the Lieb-Liniger
model can be observed in a setup with a trap.  The finite temperature
results we present below are needed for explaining the data of
Ref.~\onlinecite{Inguscio1}, see Sec.~\ref{RelToExperiment},  
and are thus elucidating the conditions realized in that experiment.

We consider non-interacting bosons at finite temperature $T$ 
in a trap of frequency $\omega_0$.
The dynamic structure factor is obtained as follows,
\begin{eqnarray}
S_N(q,\omega)&=&2\pi
\sum_{\ell,\ell'=0\atop (\ell\neq\ell')}^{\infty}
\left|
\int dx e^{iqx}\varphi^*_\ell(x)\varphi_{\ell'}(x)
\right|^2
\nonumber\\
&&\times
(1+N_\ell)N_{\ell'}\delta(\omega-\omega_{\ell\ell'}),
\label{STnonint0}
\end{eqnarray}
where $\varphi_\ell(x)$ is the harmonic oscillator state,
$N_\ell$ is the average number of particles in state $\ell$,
and
$\omega_{\ell\ell'}=(\ell-\ell')\omega_0$.
Within the grand-canonical ensemble,
$N_\ell$ obeys the Bose-Einstein distribution,
$N_\ell=\left[\exp\left((\ell\omega_0-\mu)/T\right)-1\right]^{-1}$,
with the chemical potential $\mu$
determined from the equation of state,
\begin{equation}
\sum_{\ell=0}^{\infty} N_\ell=N,
\label{NellNeqofstate}
\end{equation}
at a fixed value of $T$.
At thermal equilibrium, it is sufficient to consider
only absorption processes ($\omega>0$),
since the probabilities of energy emission and  energy absorption are
related to each other by the detailed balance equation
\begin{equation}
S(q,-\omega)=e^{-\omega/T}S(q,\omega).
\end{equation}
Assuming hereafter $\omega>0$ and evaluating
the matrix elements of $e^{iqx}$ in Eq.~(\ref{STnonint0})
on the harmonic oscillator states, we arrive at
(cf. Eq.~(\ref{Sqwnonintdirect}))
\begin{eqnarray}
\label{SNqwtempNp}
&&S_N(q,\omega)=2\pi\sum_{\ell=1}^{\infty}{\cal N}(\ell)
\frac{\left(q\lambda\right)^{2\ell}}{2^{\ell}\ell!}
e^{-\frac{q^2\lambda^2}{2}}
\delta(\omega-\ell\omega_0),\quad\quad\quad\\
&&{\cal N}(\ell)=\sum_{\nu=0}^{\infty}
\frac{\ell!\nu!}{(\ell+\nu)!}
\left[L_\nu^\ell\left(\frac{q^2\lambda^2}{2}\right)\right]^2
(1+N_{\ell+\nu})N_\nu,\quad\quad
\label{Ncalell}
\end{eqnarray}
where $L_\nu^\ell(x)$ are Laguerre polynomials.
Temperature affects the dynamic structure factor in Eq.~(\ref{SNqwtempNp})
through the shape of the function ${\cal N}(\ell)$.
At $T=0$, this function is constant, ${\cal N}(\ell)=N$,
and we recover the result of Eq.~(\ref{Sqwnonintdirect}).
In order to analyze the behavior of ${\cal N}(\ell)$ at
finite $T$, we rearrange the sum in Eq.~(\ref{Ncalell})
as follows.
We represent $(1+N_{\ell+\nu})N_\nu$ 
as the series (one may regard this series as a formal expansion
in powers of variable $\xi^\nu$)
\begin{eqnarray}
\label{Nelnup1Nnusum}
&&(1+N_{\ell+\nu})N_\nu=\sum_{p=1}^{\infty}a_p(\ell) \xi^{\nu p},\\
&&a_p(\ell)=\frac{e^{p\mu/T}\left(1-\xi^{p\ell}\right)}{1-\xi^{\ell}},
\end{eqnarray}
where $\xi=\exp\left(-\omega_0/T\right)$.
Then we carry out the summation in Eq.~(\ref{Ncalell})
for each term of the sum of Eq.~(\ref{Nelnup1Nnusum})
separately,
using the identity\cite{GradsteinRyzhik}
\begin{equation}
\sum_{\nu=0}^{\infty}\frac{\nu!L_\nu^{\ell}(x)L_\nu^{\ell}(y)z^\nu}{\Gamma(\nu+\ell+1)}=
\frac{(xyz)^{-\frac{\ell}{2}}}{1-z}e^{-z\frac{x+y}{1-z}}I_{\ell}\left(2\frac{\sqrt{xyz}}{1-z}\right),
\end{equation}
where $I_{\ell}(x)$ is the modified Bessel function of the first kind
and $|z|<1$.
As a result we obtain,
\begin{eqnarray}
\label{Ncalsel}
{\cal N}(\ell)&=&\frac{2^\ell\ell!}{\left(q\lambda\right)^{2\ell}}
e^{\frac{q^2\lambda^2}{2}}s(\ell),\\
s(\ell)&=&
\sum_{p=1}^{\infty}a_p(\ell)\frac{\xi^{-\frac{p\ell}{2}}}{1-\xi^p}
e^{-\frac{1+\xi^p}{1-\xi^p}\frac{q^2\lambda^2}{2}}
I_{\ell}\left(
\frac{
q^2\lambda^2\xi^{\frac{p}{2}}
}{
1-\xi^{p}
}
\label{sofell1}
\right).\nonumber\\
\end{eqnarray}
Further, we analyze the function $s(\ell)$;
the dynamic structure factor in Eq.~(\ref{SNqwtempNp})
is expressed through $s(\ell)$ as
\begin{equation}
S_N(q,\omega)=2\pi\sum_{\ell=1}^{\infty}
s(\ell)\delta(\omega-\ell\omega_0).
\label{SNthroughsofell}
\end{equation}
We take the continuum limit over $\ell$, assuming that 
$\omega_0$ is small, and thus, only large $\ell=\omega/\omega_0$ are of interest.
We are interested in frequencies $\omega\sim q^2/m$, which
corresponds to $\ell\sim (q\lambda)^2$.
The Bessel function in Eq.~(\ref{sofell1})
has thus to be evaluated at both large orders $\ell\gg 1$
and proportionally large values of its argument.
In order to do this, we use
Debye's asymptotic expansion of the Bessel 
function\cite{AbramowitzStegun}
\begin{eqnarray}
\label{BesselIDebye}
I_{\ell}(\ell z)&=&\frac{1}{\sqrt{2\pi\ell}}
\frac{e^{\ell\eta(z)}}{(1+z^2)^{1/4}}
\left\{1+\sum_{k=1}^{\infty}\frac{u_k(t)}{\ell^k}\right\},\quad\quad\\
\eta(z)&=&\sqrt{1+z^2}+\ln\frac{z}{1+\sqrt{1+z^2}},
\end{eqnarray}
where $u_k(t)$ are polynomials of order $3k$ of
$t=1/\sqrt{1+z^2}$.
We refer the reader to Ref.~\onlinecite{AbramowitzStegun}
for the explicit form of $u_k(t)$.
Here, we consider only the leading-order term
in Eq.~(\ref{BesselIDebye}), which amounts to 
dropping the sum over $k$.
Expanding, for consistency, also other terms in 
Eq.~(\ref{sofell1}) for large $\ell\sim (q\lambda)^2 \gg 1$, we obtain
\begin{eqnarray}
s(\ell)&=&
\frac{1}{\sqrt{\pi}q\lambda}\frac{1}{1-e^{-\ell\omega_0/T}}
\sum_{p=1}^{\infty}
\frac{1-e^{-\ell p\omega_0/T}}{\sqrt{1-e^{-2p\omega_0/T}}}
e^{p\mu/T}
\nonumber\\
&&\times
\exp\left[-\tanh\left(\frac{p\omega_0}{2T}\right)
\left(\frac{\ell}{q\lambda}-\frac{q\lambda}{2}\right)^2\right]
\label{sofellmain}
\end{eqnarray}
Equation~(\ref{sofellmain}) is our main result in this appendix.
We analyze it in greater detail below.
With the help of Eq.~(\ref{sofellmain}),
the dynamic structure factor
can be expressed,
in the continuum limit over $\omega$, as follows
\begin{equation}
S_N(q,\omega)=\frac{2\pi}{\omega_0}s\left(\omega/\omega_0\right).
\label{SNcontsofell}
\end{equation}

The convergence of the sum in  Eq.~(\ref{SNthroughsofell}) is guaranteed by the
factor $e^{p\mu/T}$ in the summand.
The ratio $\mu/T < 0$ is found for given values of $N$ and $\omega_0/T$
from Eq.~(\ref{NellNeqofstate}).
We consider here only the case $N\gg 1$.
With increasing the temperature, $\mu/T$ monotonically increases
by absolute value
from $\mu/T=-1/N$ at $T=0$ to
$\mu/T\approx -\ln(T/\omega_0N)$ at $T\gg \omega_0N$.
As a result, 
the higher the temperature the faster 
the sum in Eq.~(\ref{SNthroughsofell}) converges.
This is in contrast to the sum in Eq.~(\ref{Ncalell})
which converges fastest at low temperatures.
The presence of the factor $\tanh\left(p\omega_0/2T\right)$
in the exponent in Eq.~(\ref{Ncalell})
introduces an additional scale $p\sim T/\omega_0$, which can
be larger or smaller than the convergence scale $p\sim T/(-\mu)$.
As a result, we obtain several temperature regimes:
\begin{eqnarray}
\mbox{(i)}& T\ll\omega_0, & \mbox{low temperatures},\nonumber\\
\mbox{(ii)}& \omega_0\ll T\ll T_{*}, & \mbox{high temperatures},\nonumber\\
\mbox{(iii)}& T_{*} \ll T, & \mbox{very high temperatures},\nonumber
\end{eqnarray}
where $T_{*}$ is defined as the temperature at which $\mu=-\omega_0$.
In the limit $N\gg 1$, we obtain from Eq.~(\ref{NellNeqofstate}) 
that $T_{*}$ satisfies the transcendental equation 
$N=(T_{*}/\omega_0)\ln\left(T_{*}/\omega_0\right)$.
Thus, to leading logarithmic order ($\ln N\gg 1$), we have
$T_{*}=\omega_0 N/\ln(N)$.

\begin{figure}[t]
\includegraphics[width=0.95\columnwidth]{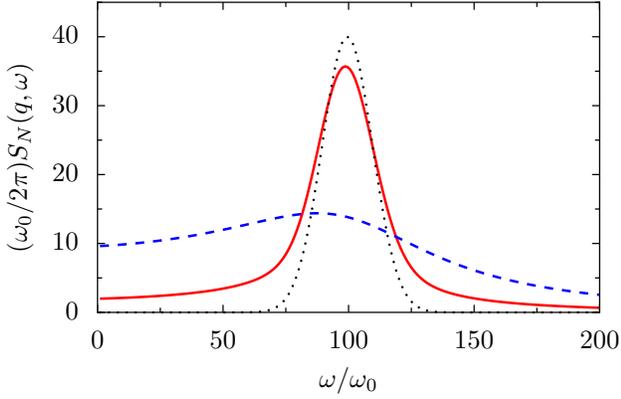}\caption{\label{sellappcfig}
Dynamic structure factor $S_{N}(q,\omega)$ of non-interacting bosons 
at different temperatures.
Dotted line:
at low temperatures, $T\ll \omega_0$, 
the factor $S_{N}(q,\omega)$ displays a Gaussian peak of width 
$\delta\omega\sim\sqrt{\omega_0q^2/m}$ centered at 
$\omega\approx q^2/2m$, see Eq.~(\ref{SqwnonintdirectGauss}).
Solid line:
at high temperatures, $\omega_0\ll T\ll \omega_0N/\ln N$,
the factor $S_{N}(q,\omega)$ retains the low-temperature peak 
at a reduced intensity and develops a much broader
(background) peak of width $\delta\omega\sim\sqrt{Tq^2/m}$
centered at the same position if $q^2/m\gg T$.
Dashed line:
at very high temperatures, $\omega_0N/\ln N\ll T$,
only the broad peak is present.
The width of this peak continues to increase
with temperature, see text at Eq.~(\ref{dtomTlogTw0N}).
The plot is made using Eqs.~(\ref{sofellmain}) and~(\ref{SNcontsofell})
for a system of $N=1000$ particles 
at temperatures $T/\omega_0=0$ (dotted), $100$ (solid), and
$250$ (dashed); a value of $(q\lambda)^2/2=100$ was used.
}
\end{figure}

Let us first consider the zero-temperature limit, $T\ll \omega_0$.
Approximating $e^{-\omega_0/T}\approx 0$ and $\mu/T\approx -1/N$ 
in Eq.~(\ref{sofellmain}),
we arrive at
\begin{equation}
s(\ell)=
\frac{1}{\sqrt{\pi}q\lambda}
\left(\sum_{p=1}^{\infty}
e^{-p/N}\right)
\exp\left[
-\left(\frac{\ell}{q\lambda}-\frac{q\lambda}{2}\right)^2\right].
\label{sofellT0}
\end{equation}
The sum in the parentheses in Eq.~(\ref{sofellT0}) equals $N$
at $N\gg 1$.
Using Eq.~(\ref{sofellT0}) in Eq.~(\ref{SNcontsofell}),
we recover the zero-temperature result of 
Eq.~(\ref{SqwnonintdirectGauss}), up to  
$\omega/\omega_0\leftrightarrow (q\lambda)^2/2$
in the prefactor, which is within the accuracy of
the asymptotic expansion.

In the high temperature regime, $\omega_0\ll T\ll T_{*}$,
we split the sum in Eq.~(\ref{sofellmain}) into two parts.
In the first part, the summation index $p$ runs from
$p=1$ to $p\sim T/\omega_0$.
Here, we approximate 
$\tanh\left(p\omega_0/2T\right)\approx p\omega_0/2T$
and replace the sum over $p$ by an integral.
In the second part, the summation can be carried out
explicitly as in the case of Eq.~(\ref{sofellT0}),
but with the lower summation bound at $p\sim T/\omega_0$.
As a result, we obtain 
\begin{eqnarray}
\label{sofellT12}
s(\ell)&=& s_1(\ell)+s_2(\ell),\\
s_1(\ell)&=&
\frac{1}{\sqrt{\pi}q\lambda}\frac{T/\omega_0}{1-e^{-\ell\omega_0/T}}
\int_{0}^{c}dz
\frac{1-e^{-\ell z}}{\sqrt{1-e^{-2 z}}}
e^{\mu z/\omega_0}
\nonumber\\
&&\times
\exp\left[-\frac{z}{2}
\left(\frac{\ell}{q\lambda}-\frac{q\lambda}{2}\right)^2\right],
\label{sofellT1}
\\
s_2(\ell)&=&
\frac{1}{\sqrt{\pi}q\lambda}
\frac{e^{-(\ell/q\lambda-q\lambda/2)^2}}{1-e^{-\ell\omega_0/T}}
\frac{e^{c\mu/\omega_0}}{1-e^{\mu/T}},
\label{sofellT2}
\end{eqnarray}
where $c$ is a number order one, which can be fixed by
requiring $\partial s(\ell)/\partial c =0$.
In practice, choosing a fixed value for $c$ 
in the range between $1$ and $5$ yields accurate results.
The integral in Eq.~(\ref{sofellT1}) can be expressed
through special functions if desired.
The qualitative behavior of the functions $s_1(\ell)$
and $s_2(\ell)$, in the considered temperature regime, is as follows.
The function $s_2(\ell)$ is, up to a prefactor, the zero-temperature
result, see Eq.~(\ref{SqwnonintdirectGauss}).
It contributes to $S_N(q,\omega)$ with a relatively
sharp peak at $\omega\approx q^2/2m$.
In contrast, the function $s_1(\ell)$ has a much broader
peak centered at the same place as the peak of $s_2(\ell)$,
provided $q^2/2m \gg T$.
The dynamic structure factor 
$S_N(q,\omega)$ is, therefore, a sharp peak with broad tails,
see solid line in Fig.~\ref{sellappcfig}.
As the temperature increases, the relative spectral weight shifts
from the peak to the tails.
The width of the sharp peak is given by Eq.~(\ref{width1dq})
with $\delta q\sim 1/\lambda$.
The width of the broad peak is given by the same equation but
with $\delta q\sim \sqrt{m T}$.
The physical explanation for the coexistence of the sharp and
broad peaks is the fact that in this temperature regime
a macroscopic fraction of particles occupy the ground state of the 
harmonic potential. 
The sharp peak is due to the particles in the ground state
and the broad peak is due to the particles in the thermal tail 
of the Bose-Einstein distribution function.

In the ``very high'' temperature regime, $T\gg T_*$,
the contribution $s_{2}(\ell)$ can be neglected.
Furthermore, the upper bound of integration
in Eq.~(\ref{sofellT1}) can be extended to infinity,
since the factor $e^{\mu z/\omega_0}$ 
suppresses the integrand before the upper bound
is reached.
As a result, we obtain
\begin{eqnarray}
s(\ell)&=&
\frac{1}{q\lambda}\frac{T/2\omega_0}{1-e^{-\ell\omega_0/T}}
\left[
\frac{\Gamma(\mu_-)}{\Gamma\left(\frac{1}{2}+\mu_-\right)}-
\frac{\Gamma(\mu_+)}{\Gamma\left(\frac{1}{2}+\mu_+\right)}
\right],
\nonumber\\
\mu_\pm&=&
\frac{-\mu}{2\omega_0}+\frac{1}{4}
\left(\frac{\ell}{q\lambda} \pm \frac{q\lambda}{2}\right)^2.
\label{sofellTTT}
\end{eqnarray}
If $q^2/2m\gg T$, then the dynamic response has a peak
at $\omega\approx q^2/2m$ with width
$\delta\omega\sim (q/m)\sqrt{(-\mu) m}$.
A simple expression for $\mu$ can be given only
if temperature is sufficiently high, $T\gg \omega_0 N$, 
see text below Eq.~(\ref{SNcontsofell}).
Then, the width of the peak is
\begin{equation}
\delta\omega\sim q\sqrt{\frac{T}{m}\ln\left(\frac{T}{\omega_0N}\right)}.
\label{dtomTlogTw0N}
\end{equation}
If $q^2/2m\ll T$, then $S_N(q,\omega)$ is a monotonically
decaying function of $\omega$ on the scale of temperature.

We show the behavior of $S_N(q,\omega)$ in different temperature
regimes in Fig.~\ref{sellappcfig}.


\begin{references}
\bibitem{footnote1} 
Accidentally, the square-root dependence in
  Eq.~(\ref{tgm}), derived analytically for small $J/U$ in
  the limit of smooth confinement, agrees quite well with the numerical
  simulation of Ref.~\onlinecite{Rigol08}. 
  The numerically found coefficient is approximately $2.7$ 
  in place of $8/\pi\approx 2.54$ in Eq.~(\ref{tgm}).
\bibitem{noteCMM}
Strictly speaking, it is the geometrical center that we are using; 
however, we call it center of mass to draw analogy to the standard
two-body problem.
\bibitem{note25}
Invoking averaging over an ensemble of 1D systems, 
see Appendix~\ref{AppAvOverTubes},
we verify that the 
definition\cite{Weiss,Inguscio1} 
$\gamma_{\rm av}:=\int dN P(N)N\gamma(N)/\int dN P(N)N$,
with $P(N)$ in Eq.~(\ref{PofM}),
yields the value of $\gamma_{\rm av}\approx 0.54$.
We also note that the experiment is carried out in the
regime $mg^2 \ll \omega_0N_{\rm max}$, for which we find
that $\gamma_{\rm av}=(5/3)\gamma$.
\bibitem{note2}
For the CIC state in Fig.~\ref{motttrapfig}
at $q=\pi/a$, the response of the Mott phase is 
$(3/4)(N/N_0)^2$ times larger than the response of the compressible caps.
\bibitem{note3}
Note that the expression for $N_{i,j}$ in Ref.~\onlinecite{Bloch}
apparently contains a misprint, $5N/2\pi N_{0,0}\to 2\pi N_{0,0}/5N$.
\end{references}
\end{document}